\documentclass{aa} %Two-Column
\usepackage{amsmath,amsfonts,latexsym,amscd, tabularx}
\usepackage[varg]{txfonts}
\usepackage{graphicx}
\usepackage{gensymb}
\usepackage{longtable}
\usepackage{natbib, url}
\begin{document}

\title{A New Radio Molecular Line Survey of Planetary Nebulae:}
\subtitle{HNC/HCN as a Diagnostic of Ultraviolet Irradiation}

\author{J. Bublitz\inst{1,2}
\and J. H. Kastner\inst{2}
\and M. Santander-Garc\'{i}a\inst{3}
\and V. Bujarrabal\inst{3}
\and J. Alcolea\inst{3}
\and R. Montez Jr.\inst{4}
}

\institute{Institut de Plan\'{e}tologie et d'Astrophysique de Grenoble (IPAG) UMR 5274, F-38041, Grenoble, France
\and Center for Imaging Science and Laboratory for Multiwavelength Astrophysics, Rochester Institute of Technology, 54 Lomb Memorial Drive, Rochester, NY 14623, USA
\and Observatorio Astron\'{o}mico Nacional, Alfonso XII, 3, 28014, Madrid, Spain
\and Smithsonian Astrophysical Observatory, 60 Garden Street, Cambridge, MA 02138, USA
}

\date{Received <date> /
Accepted <date>} 

%%%%%%%%%%%%%%%%%%%% Abstract  %%%%%%%%%%%%%%%%%%%%
\abstract{
Certain planetary nebulae contain shells, filaments, or globules of cold gas and dust whose heating and chemistry are likely driven by UV and X-ray emission from their central stars and from wind-collision-generated shocks. We present the results of a survey of molecular line emission in the 88-236 GHz range from nine nearby ($<$1.5 kpc) planetary nebulae spanning a range of UV and X-ray luminosities, using the 30 m telescope of the Institut de Radioastronomie Millim\'{e}trique. Rotational transitions of thirteen molecules, including CO isotopologues and chemically important trace species, were observed and the results compared with and augmented by previous studies of molecular gas in PNe. Lines of the molecules HCO$^+$, HNC, HCN, and CN, which were detected in most objects, represent new detections for five planetary nebulae in our study. Specifically, we present the first detections of $^{13}$CO (1-0, 2-1), HCO$^+$, CN, HCN, and HNC in NGC 6445; HCO$^+$ in BD+30$\degree$3639; $^{13}$CO (2-1), CN, HCN, and HNC in NGC 6853; and $^{13}$CO (2-1) and CN in NGC 6772. Flux ratios were analyzed to identify correlations between the central star and/or nebular UV and X-ray luminosities and the molecular chemistries of the nebulae. This analysis reveals a surprisingly robust dependence of the HNC/HCN line ratio on PN central star UV luminosity. There exists no such clear correlation between PN X-rays and various diagnostics of PN molecular chemistry. The correlation between HNC/HCN ratio and central star UV luminosity demonstrates the potential of molecular emission line studies of PNe for improving our understanding of the role that high-energy radiation plays in the heating and chemistry of photodissociation regions.
}

\maketitle

%\tableofcontents
%\newpage
%\renewcommand{\thepage}{\arabic{page}}% Arabic numerals for page counter
%\setcounter{page}{1}% Start page number with 1
%\listoftodos[Notes]

%%%%%%%%%%%%%%% INTRODUCTION %%%%%%%%%%%%%%%
\section{Introduction} \label{Intro}
%background info on PNe community and molecule analysis

Planetary nebulae (PNe) arise from outflowing stellar mass during the late evolutionary stages of intermediate-mass stars ($\sim$0.8-8.0 M$_\odot$). These stars represent a significant stellar population in the Galaxy \citep{Blocker01, Edwards14}. They progress through the main sequence and red giant branch (RGB) evolutionary phases, where core H and He fusion occur respectively, into the shell-burning asymptotic giant branch (AGB) phase. Once the star arrives there, slow AGB winds, originating in shocks and pulsations and driven by radiation pressure on dust, remove the bulk of the stellar envelope at mass loss rates of 10$^{-7}$-10$^{-4}$ M$_{\odot}$/yr \citep{Zack13, Blocker01}. The presence of a close companion star can accelerate and otherwise profoundly affect this mass-loss process and the evolution of the resulting envelope \citep[e.g.,][and references therein]{DeMarco17}.

When the AGB envelope depletes to a mass of $\sim$10$^{-2}$ M$_\odot$, mass loss ceases and the star progresses to the post-AGB, and soon after, it becomes the central star of a PN (CSPN). Here, the cold, dusty AGB envelope is suddenly exposed to the hot (30-200 kK) post-fusion core's ionizing UV and (in some cases) X-ray emission, which photodissociates and ionizes the envelope gas. This newly ionized gas constitutes the PN. Fast winds from the pre-white dwarf star (or its companion) may also slam into the slower moving ejected envelope, generating shocks that heat plasma to temperatures exceeding $\sim$10$^6$ K \citep[and references therein]{Kastner12}. Hot bubbles that form from this wind interaction have been found to be X-ray-luminous, thereby producing another ionization mechanism within the PN. Interactions from the winds that form the bubbles are also thought to continue shaping the nebula throughout its lifetime \citep{Balick02, HE12}. Atomic gas and dust within the nebula limit the penetration depth of incident UV photons from the CSPN, leaving the outer layers of the nebula insulated from them. Only higher energy (X-ray) photons, $>$0.5 keV, can penetrate the dense, neutral gas of the PN to ionize the cold, molecule-rich outer shells \citep{Tielens85}.

Millimeter CO and infrared H$_2$ lines gave the first view of PN molecular gas 3 decades ago and have remained commonplace probes into the shells of ejected mass since \citep[and references therein]{Zuckerman88, Huggins89, Bachiller91, Huggins96}. Due to its high abundance, low excitation requirements and the low critical densities of its rotational transitions, CO is the most commonly observed and more widely reliable molecular species found in PNe \citep{Huggins96}. Infrared lines of H$_2$ are also observed from PNe, with detections of H$_2$ mainly confined to bipolar (axisymmetric) nebulae \citep[and references therein]{Kastner96, Zuckerman88}. %The presence of vibrationally excited H$_2$ is generally attributed to the presence of shocked gas, as it requires temperatures at $>$10$^3$ K to become populated \citep{Zuckerman88}.
The complex interplay of PN central star radiation and composition of the proto-PN produce a rich environment for molecular chemistry that has served as motivation for various radio molecular line surveys \citep[e.g.,][]{Bachiller97, Zhang08, Edwards14, Schmidt18}. Whereas the foregoing surveys were largely restricted to well-studied objects, the recent molecular line surveys of \cite{Schmidt16, Schmidt17, Schmidt17b} extended the PN sample coverage to younger objects.

In this paper we present molecular line surveys of nine PNe, obtained with the Institut de Radioastronomie Millim\'{e}trique 30 m telescope within the frequency range 88-236 GHz. These observations were specifically intended to explore the utility of potential tracers of high-energy irradiation of molecular gas. We report new detections of molecules and/or molecular transitions in five of these PNe, and place these results in the context of previously reported PN molecular line detections and measurements. We then evaluate the integrated flux ratios of observed molecular lines to study correlations between the molecular chemistry and high-energy radiation properties of the CSPNe. Specifically, we strive to find tracers of non-LTE chemistry due to X-irradiation of molecular gas and the effects of CSPN UV emission on the photodissociation regions (PDR) within PNe.

%%%%%%%%%%%%%%% Observations %%%%%%%%%%%%%%%
\section{Observations and Data Reduction}
%basic overview of PNe observed and instrument stats
%Scans were averaged, baselines subtracted, beam temperature adjusted

%	\input{/Users/jtb1435/Documents/School_Work/Research/Observations/XPNE_Survey/Tables/Table1Alt1.tex} %T1A1
%	\input{/Users/jtb1435/Documents/School_Work/Research/Observations/XPNE_Survey/Tables/Table1Alt2.tex} %T1A2

%%%%%%%%%%%%%%%%%%% Summary Table Part 1%%%%%%%%%%%%%%%%%%%\
\begin{table*}
	\begin{center}
	\caption{Summary of Physical Data for Observed PNe and their Central Stars}
	\noindent\makebox[\textwidth]{
		\label{T1A1}
		{\scriptsize
		\begin{tabular}{lcccccccccc}
	\hline 
	\hline
	Name & Morphology$^1$ & Angular & D    & R   & Age & Ionized PN & CSPN & $T_\star$ & Mass$_{CS}$ & Refs.$^3$ \\
		& (F08/SMV11)	& Radius$^2$ ($^{\prime \prime}$)  &(kpc)&(pc) & (10$^3$ yr) &  Mass (M$_{\odot}$) & sp type & (kK) & (M$_{\odot}$)  & \\
	\hline	
	%			Morphology		AngRad 	      D	R	  Age	  M_PN   Spec  	   T*  MassCS    Refs
	BD+303639 	&  Er/Ecsarh		& 4   		& 1.52 & 0.02 &  0.8	& 0.01 & [WC9]	& 32   & 0.58	& a, b, c\\
	NGC 7027 	& Bs/Mctspih		& 14		& 0.92 & 0.03 &  0.7	& 0.05 & ...	& 175 & 0.67	& c, d \\
	NGC 6445 	& Bs/Mpi			& 17 		& 1.38 & 0.14 &  5.3	& 0.18 & ...	& 170 & 0.64	& e, f, c \\
	NGC 7008 	&  Efp/Bs			& 43 		& 0.97 & 0.15 &  5.7	& 0.08 & O(H)	& 97   & 0.55	& a, g, c \\
	NGC 6720 (M57) & Ebmr(h)/Ecsh	& 35 		& 0.70 & 0.13 &  7.0	& 0.09 & hgO(H)& 112 & 0.66	& a, h \\
	NGC 6853 (M27) & Ebm(h)/Bbpih	& 49 		& 0.38 & 0.37 &  10.0 & 0.41 & DAO	& 114 & 0.63	& a, i, j \\
	NGC 6772 	& Ep/E			& 32 		& 1.27 & 0.22 &  10.9 & 0.17 & ...	& 135 & 0.64	& e, k, c \\
	NGC 7293 	&  Bams(h)/Ltspir	& 402 	& 0.20 & 0.46 &  16.3 & 0.35 & DAO	& 107 & 0.63	& i, k, a \\
	NGC 6781 	& Bam(h:)/Bth		& 53 		& 0.95 & 0.32 &  20.0 & 0.44 & DAO	& 112 & 0.57	& a, l, c \\
	\hline
			\end{tabular}
	    } }
	 \vspace{1mm}
	 \end{center}
	\tablefoot{\\
	$^{1.}$ Morphology descriptions as defined in \cite{Frew08}(F08); a: asymmetry present, B: bipolar, b: bipolar core, E: elliptical, f: filled amorphous center, (h): distinct outer halo, m: multiple shells, p: point symmetry present, r: dominant ring structure, s: internal structure. \\
		Morphology description modified slightly from \cite{Sahai11}(SMV11); a: ansae, B: bipolar, c: closed outer lobes, E: elongated, h: halo, i: inner bubble, L: collimated lobe pair, M: multipolar, p: point symmetry, r: radial rays, s: CSPN apparent, t: bright central toroidal structure. \\
	$^{2.}$ Angular radii estimated using DSS data.\\
	$^{3.}$ References for PN data: 
	a: \cite{Frew16}, 
	b: \cite{Li02}, 
	c: \cite{Kastner12}, 
	d: \cite{Latter00}, 
	e: \cite{Stanghellini08}, 
	f: \cite{Phillips84}, 
	g: \cite{Gorny97}, 
	h: \cite{ODell07}, 
	i: \cite{Gaia}, 
	j: \cite{ODell02}, 
	k: \cite{Ali12}, 
	l: \cite{Ueta14} \\
	   }
	 \end{table*}
	
%%%%%%%%%%%%%%%%%%% Summary Table Part 2%%%%%%%%%%%%%%%%%%%\
\begin{table*}
	\begin{center}
	\caption{Radiative Properties of PNe$^1$}
	\label{T1A2}
	\begin{tabular}{lcccc}
	\hline
	\hline
	Name & L$_{UV}$ & X-Ray & L$_X$ & E$_{median}$ \\
		 & log(ergs s$^{-1}$) & Source$^2$ & log(ergs s$^{-1}$) &(keV) \\
	\hline
	%				Luv		X	Lx	     E
	BD+303639		& 37.23 &	D & 32.20 & 0.74 \\
	NGC 7027		& 37.47 &	D & 32.11 & 0.97  \\
	NGC 6445		& 36.57 &	P & 30.10 & 1.04  \\
	NGC 7008		& 36.72 &	P & 29.48 & 0.85  \\
	NGC 6720		& 36.18 &	N & $\leq$28.61  & ...  \\
	NGC 6853		& 36.09 &	P & 29.15  & 0.18  \\
	NGC 6772		& 36.01 &	N & $\leq$28.90 & ...  \\
	NGC 7293		& 35.55 &	P & 29.94 & 0.89  \\
	NGC 6781		& 36.15 &	N & $\leq$28.70 & ...  \\
	\hline
	\end{tabular}
	 \end{center}
\tablefoot{
	   \\
	   $^1$ PN data obtained from \cite{Montez15} unless otherwise specified. X-Ray sources and median X-ray photon energies (E$_{median}$) from \cite{Kastner12}, except NGC 7027 and BD+30 from \cite{Montez15}. Upper limit X-ray luminosities for non-detections by Montez Jr. (unpublished). \\
	   $^2$ P = point-like, D = diffuse, N = non-detection
	   }
\end{table*}

	\begin{figure*}
	\begin{tabular}{@{}ccc@{}}
	\includegraphics[width=0.33\linewidth]{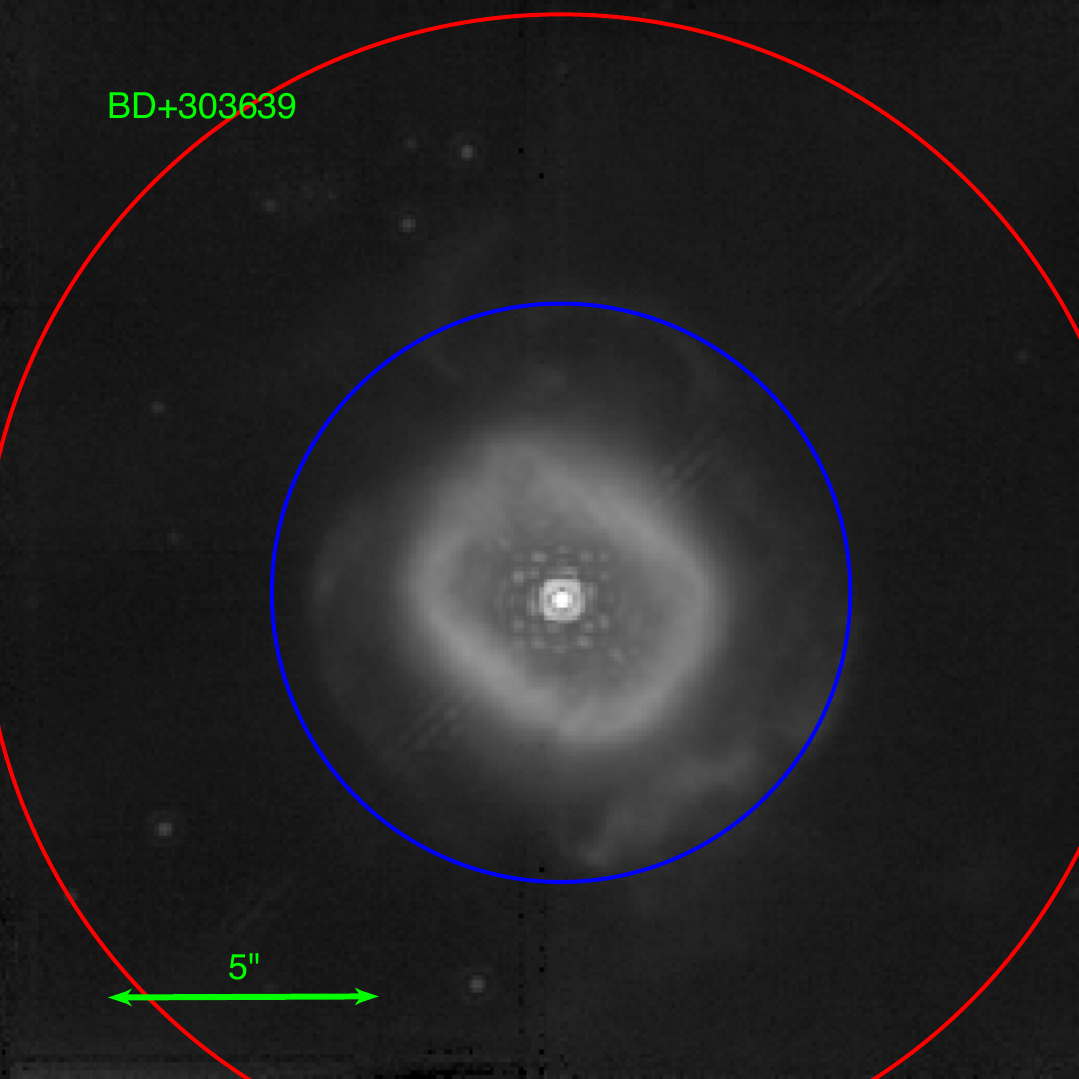} &
	\includegraphics[width=0.33\linewidth]{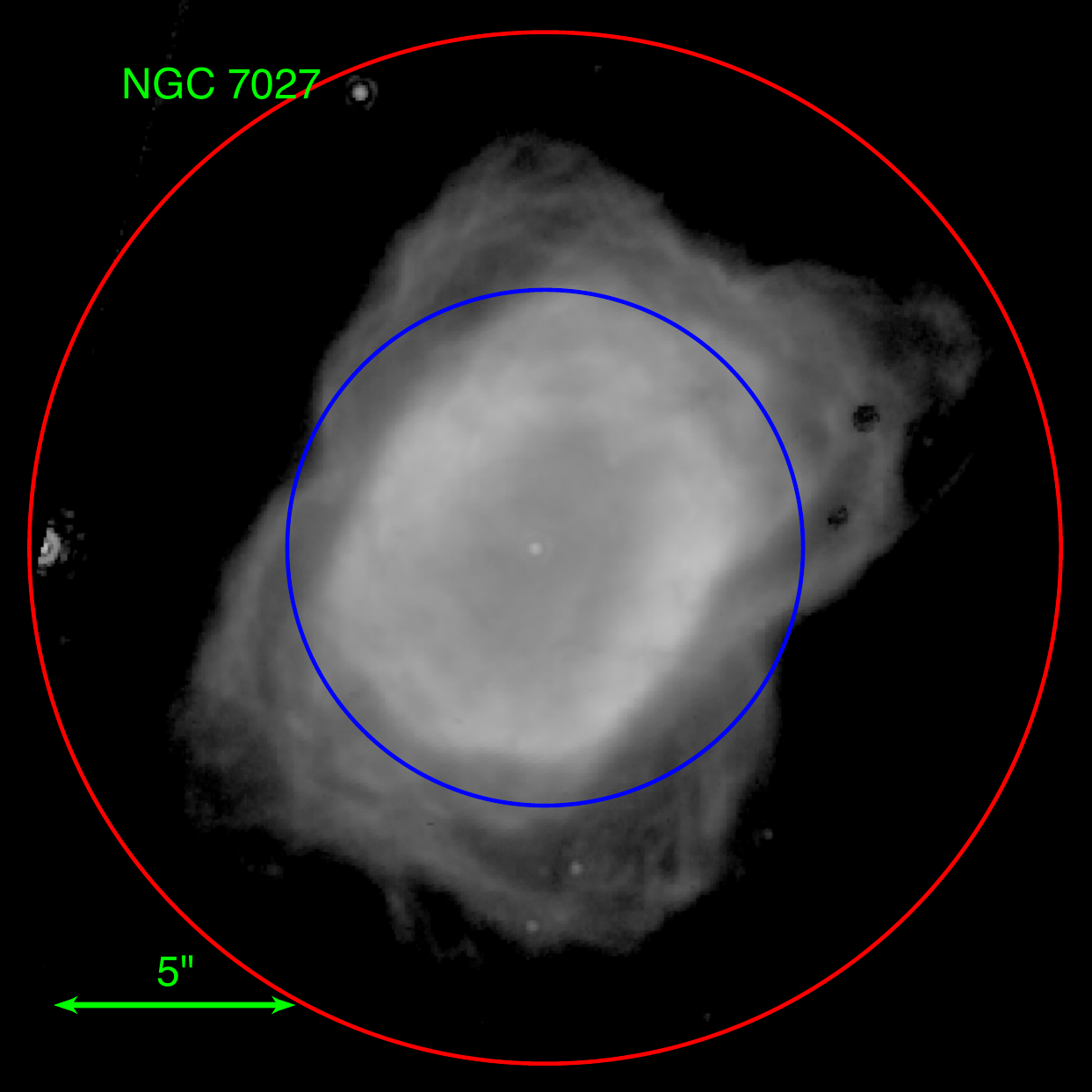} &
	\includegraphics[width=0.33\linewidth]{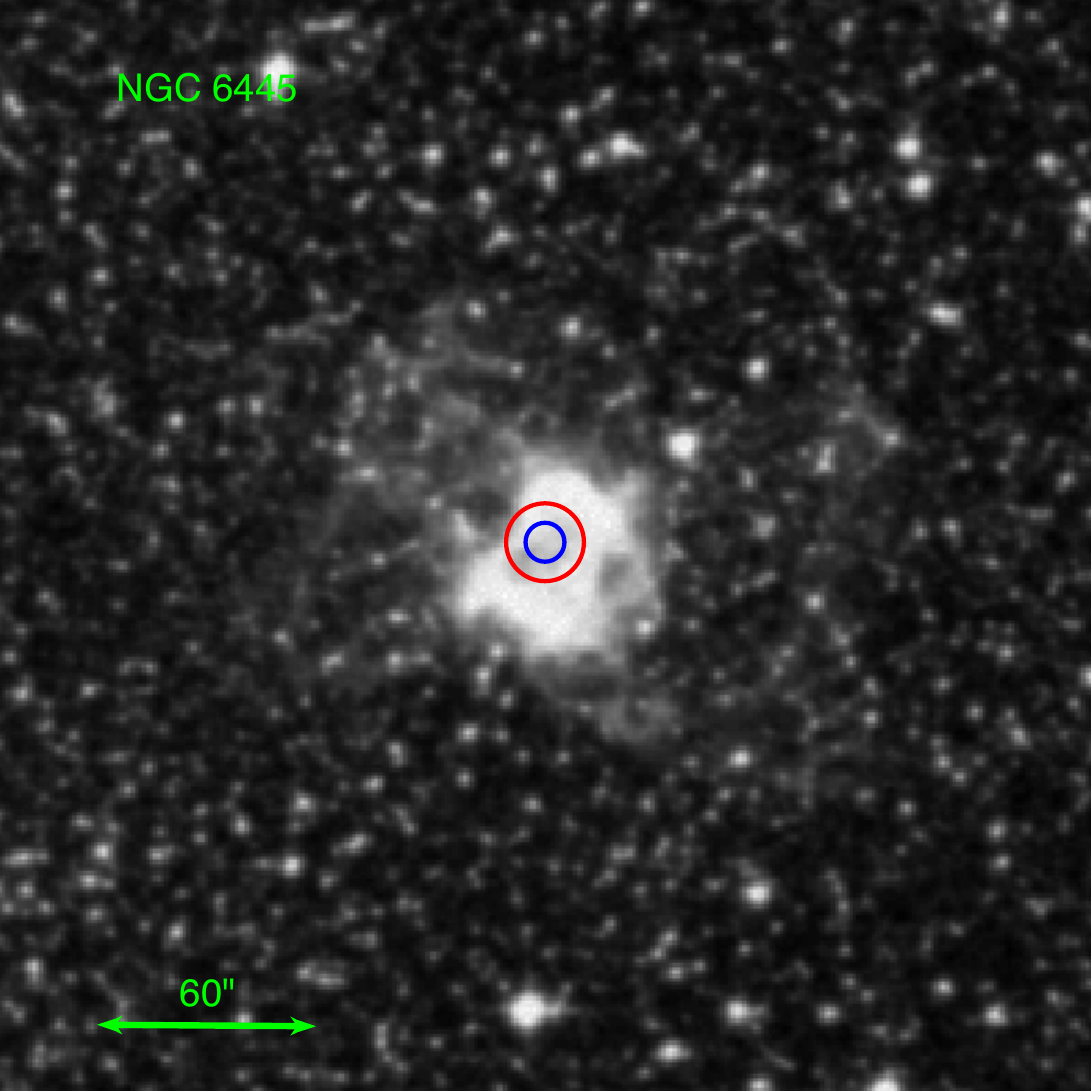} \\
	\includegraphics[width=0.33\linewidth]{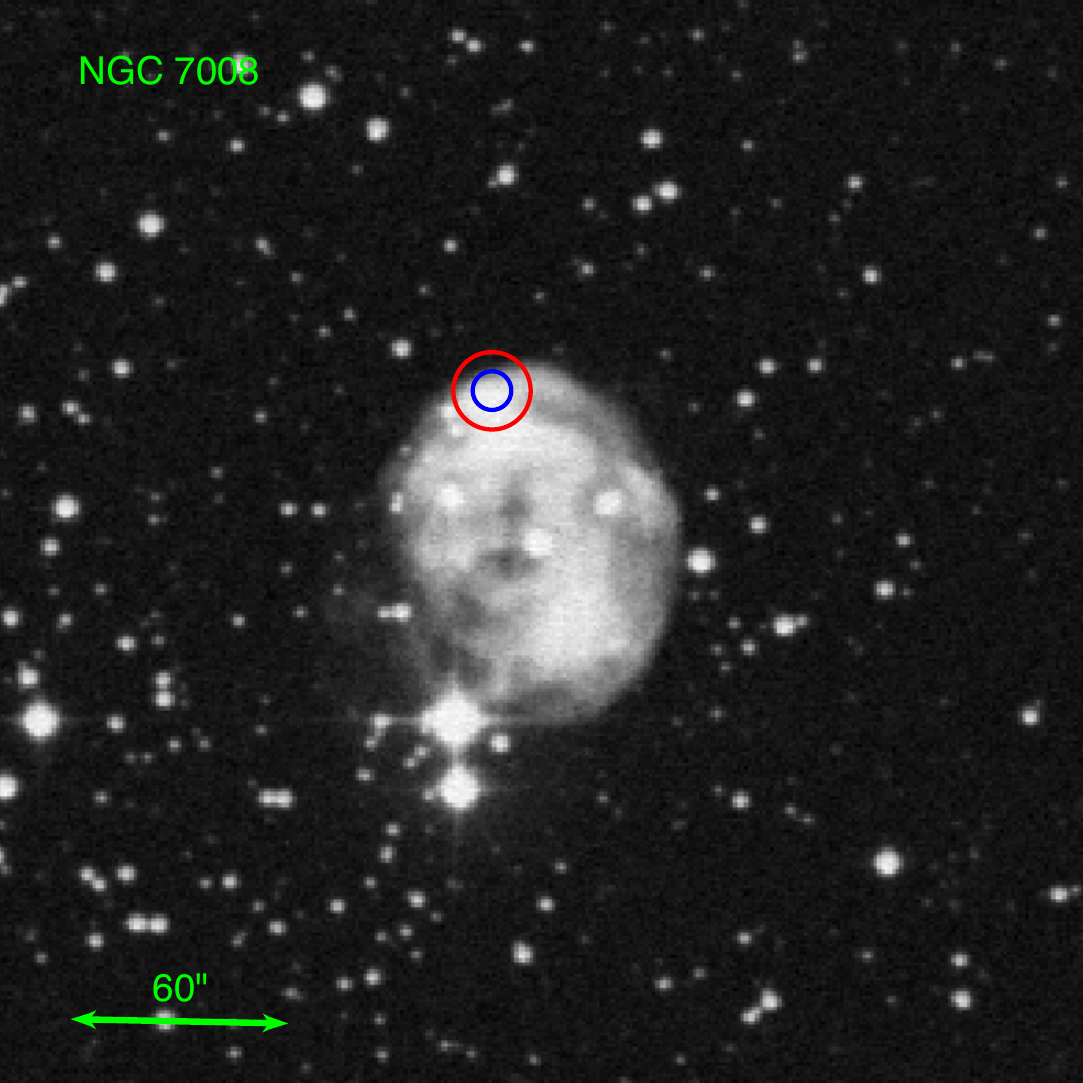} &
	\includegraphics[width=0.33\linewidth]{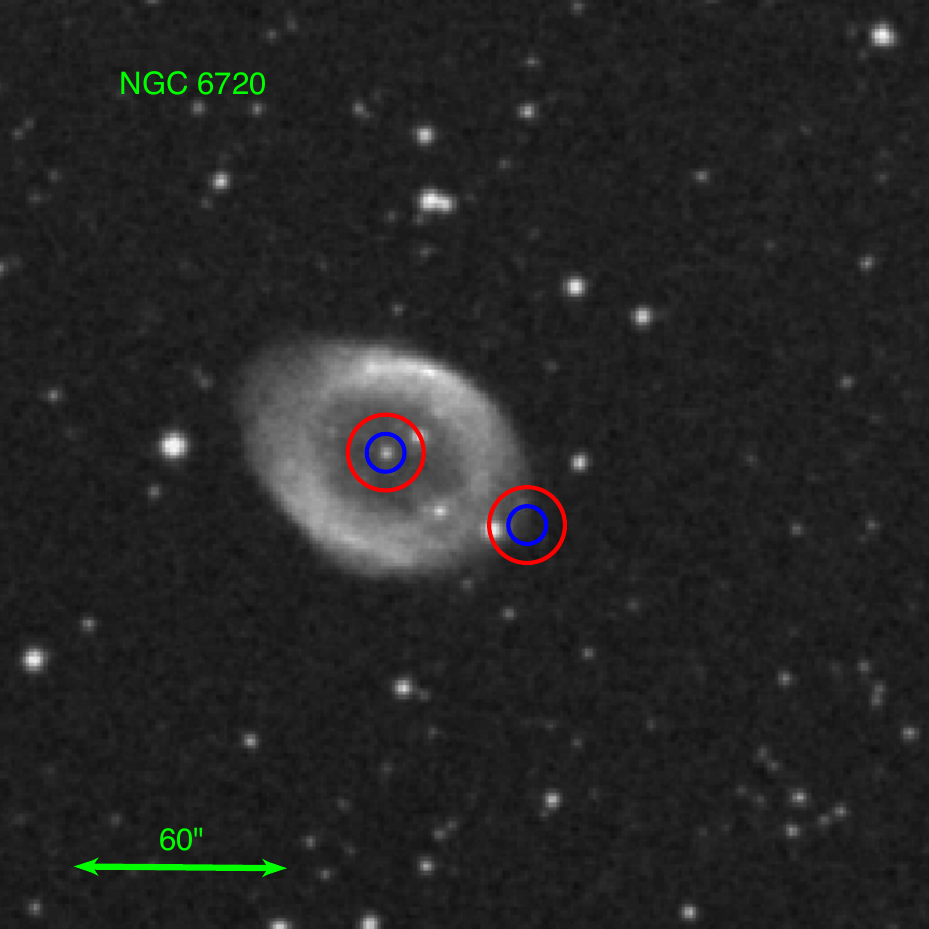} &
	\includegraphics[width=0.33\linewidth]{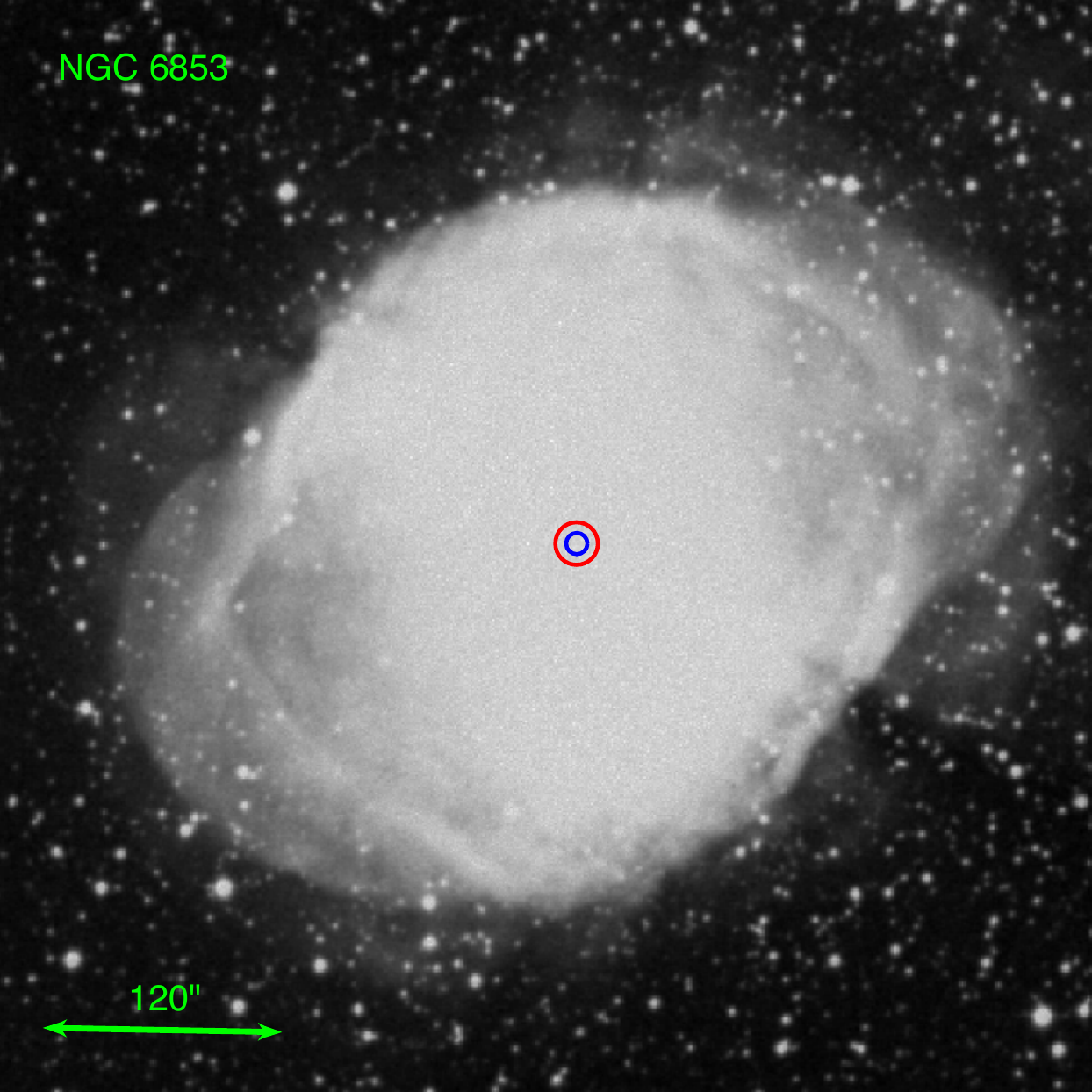} \\
	\includegraphics[width=0.33\linewidth]{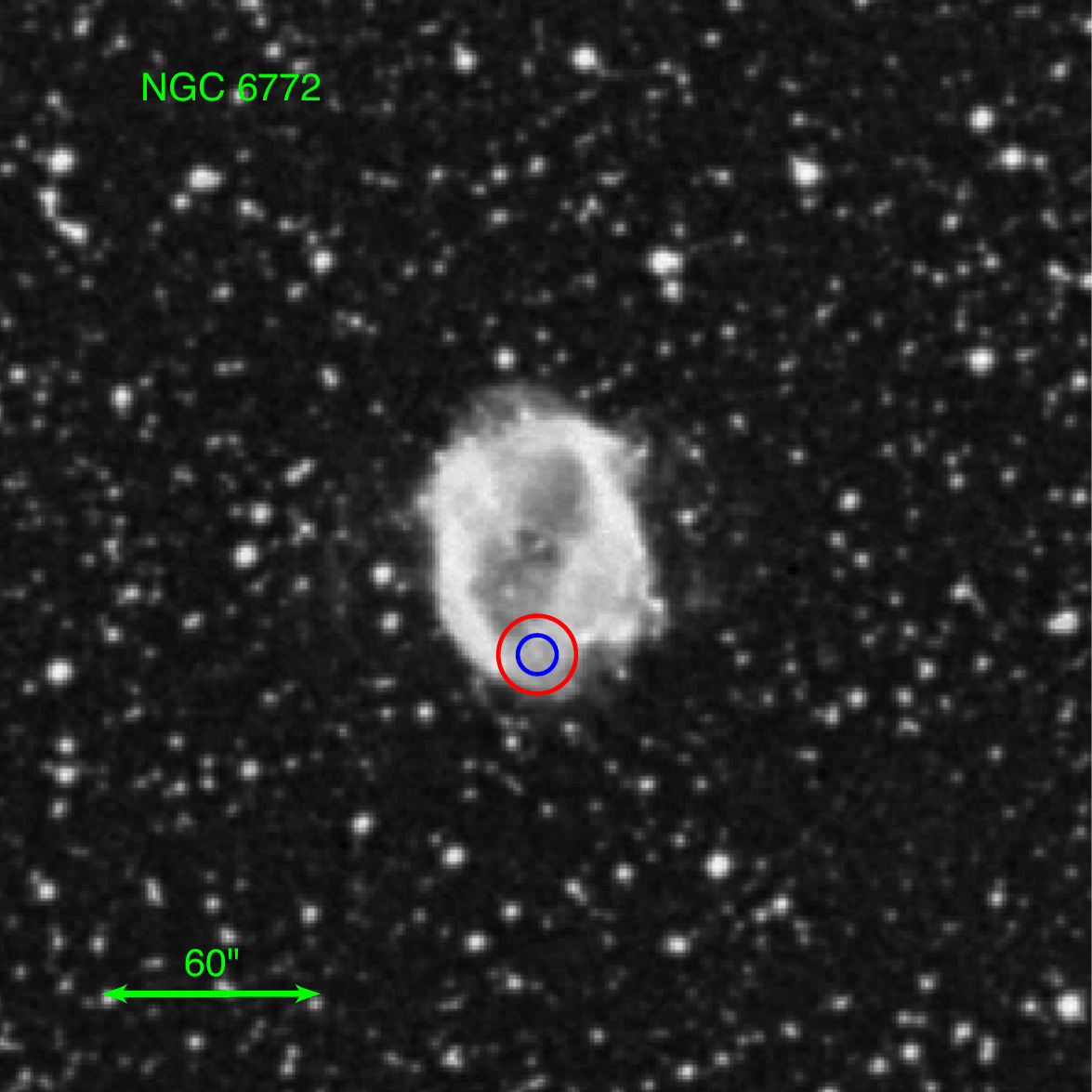} &
	\includegraphics[width=0.33\linewidth]{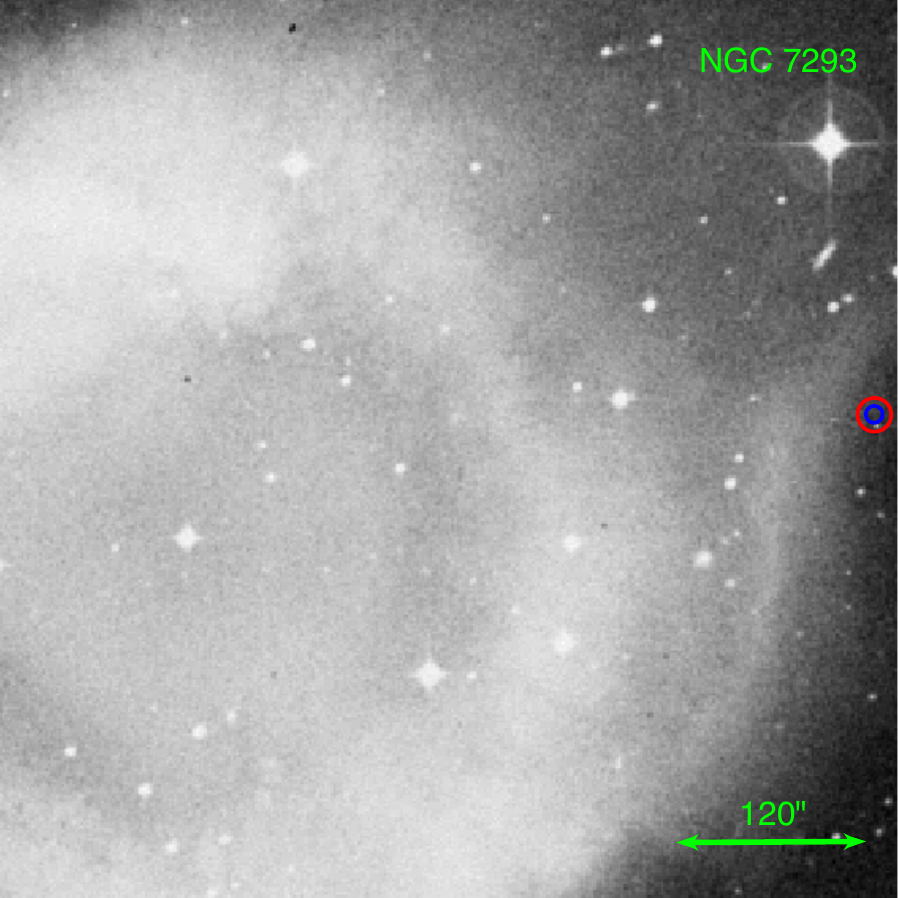} &
	\includegraphics[width=0.33\linewidth]{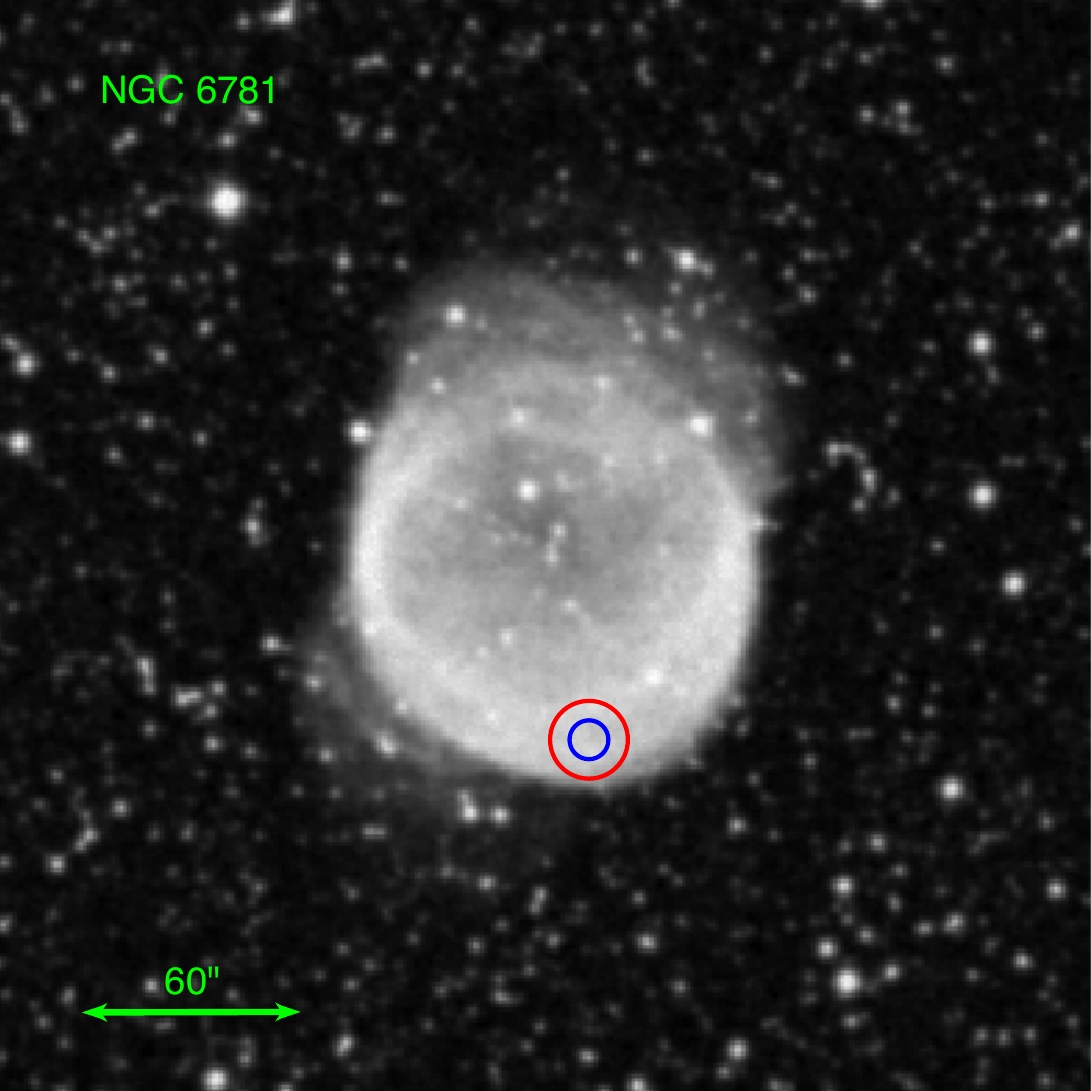} \\
	\end{tabular}
	\caption{POSS II red images for seven of the nine PNe observed in our study, with HST NICMOS images for BD+30 and NGC 7027. Circles illustrate beam diameter (half-power beam width) of the telescope and the position observed, where $^{12}$CO (1-0) and $^{12}$CO (2-1) were chosen to represent the range in beam size across frequency (red circle and blue circle, respectively). North is up, East is to the left in each frame.}
	\label{PNePositions}
\end{figure*}

\indent The sample of nine PNe studied here were selected from the larger Chandra Planetary Nebula Survey PN sample \citep{Kastner12, Freeman14}. The ChanPlaNs survey targeted 35 objects within 1.5 kpc of Earth with the aim to identify and categorize PN X-ray emission, so as to understand the mechanisms giving rise to X-rays within PNe (Kastner et al. 2012). The ChanPlaNS survey mostly included high-excitation PNe and yielded X-ray detections from central star (point-like) and nebular (diffuse) X-ray sources.

We selected those ChanPlaNS PNe accessible to the IRAM 30 m telescope that had previously been detected\footnote{The lone exception, NGC 7008, was included in this survey due to its large X-ray luminosity.} in CO and/or H$_2$. Physical properties of these nine PNe and their central stars are listed in Table \ref{T1A1}. These properties include the morphology, angular radius in arcsec, distance, average radius, age, and ionized mass of each PN, along with the spectral type, temperature, and mass of the CSPN as compiled from the literature. The ultraviolet luminosity, X-ray luminosity, median energy of emission, and X-ray source properties for each of the planetaries are listed in Table \ref{T1A2}.

Data were collected for the nine listed PNe during an observing run on the IRAM 30 m telescope on Pico Veleta from 5th to 10th June, 2012. The positions for each targeted PN are listed in Table \ref{Positions} and illustrated in Figure \ref{PNePositions}. The observations were either centered on the PN central stars or the nebular rims; we obtained observations at two positions within NGC 6720, one toward the central star and one toward the rim. Frequencies in the regime of 88 GHz to 236 GHz were chosen to target key molecules for tracing chemistry. The molecular transitions detected during the course of the survey are listed in Table \ref{Transitions}.

\begin{table*}
	\begin{center}
	\caption{Observed Positions and Integration Times}
	\label{Positions}
	\begin{tabular}{l c c c}
	\hline
	\hline
	Source & $\alpha$ (J2000.0) & $\delta$ (J2000.0) & Representative \\
	 & & & Integration Time (hrs) \\ %(sec)
	\hline
	BD+303639 & 19$^h$34$^m$45.23$^s$ & 30$\degree$30$^{\prime}$58.90$^{\prime \prime}$	& 5.76 \\ %2.075$\times$10$^4$ \\
	NGC 7027 & 21$^h$07$^m$1.59$^s$ & 42$\degree$14$^{\prime}$10.20$^{\prime \prime}$	& 2.56 \\ %9.233$\times$10$^3$ \\
	NGC 6445 & 17$^h$49$^m$15.21$^s$ & -20$\degree$00$^{\prime}$34.50$^{\prime \prime}$	& 6.98 \\ %2.512$\times$10$^4$ \\
	NGC 7008 & 21$^h$00$^m$34.36$^s$ & 54$\degree$33$^{\prime}$17.20$^{\prime \prime}$ 	& 4.11 \\ %1.481$\times$10$^4$ \\
	NGC 6720 & 18$^h$53$^m$35.08$^s$ & 33$\degree$01$^{\prime}$45.03$^{\prime \prime}$ 	& 2.98 \\ %1.074$\times$10$^4$ \\	
	NGC 6720 Rim & 18$^h$53$^m$31.90$^s$ & 33$\degree$01$^{\prime}$25.00$^{\prime \prime}$  & 2.98 \\ %1.074$\times$10$^4$ \\
	NGC 6853 & 19$^h$59$^m$36.38$^s$ & 22$\degree$43$^{\prime}$15.80$^{\prime \prime}$ 	& 11.01 \\ %3.963$\times$10$^4$ \\
	NGC 6772 & 19$^h$14$^m$36.37$^s$ & -2$\degree$42$^{\prime}$55.00$^{\prime \prime}$  	& 6.89 \\ %2.482$\times$10$^4$ \\
	NGC 7293 & 22$^h$29$^m$7.50$^s$ & -20$\degree$48$^{\prime}$58.60$^{\prime \prime}$ 	& 7.79 \\ %2.803$\times$10$^4$ \\
	NGC 6781 & 19$^h$18$^m$27.42$^s$ & 6$\degree$31$^{\prime}$29.30$^{\prime \prime}$ 	& 5.43 \\ %1.953$\times$10$^4$ \\
	\hline
	\end{tabular}
	\end{center}
	\tablefoot{\\ Integration times sampled from HCN, HCO$^+$, and HNC spectral region observations for each PN.
	}
\end{table*}

We used the Eight MIxer Receiver (EMIR), which is able to observe simultaneously in the 1~mm and 3~mm bands. Dual polarization mode was used whenever each specific frequency setting made it possible, and the lines from the two polarizations were averaged after checking that their intensities were compatible. We recorded the data with the FTS200 backend, providing a resolution of 200 kHz, which translates into velocity resolutions of $\sim$0.5 and $\sim$0.25 km~s$^{-1}$ in the 88 and 236 GHz regimes, respectively, and a sufficient coverage of the lines listed above.

The beam sizes (Half Power Beam Widths) and beam efficiencies for the telescope ranged from 10.4" and 0.60, respectively, for the CO$^+$ lines at 236 GHz to 27.8" and 0.80, respectively, for the HCN line at 88.6 GHz. Saturn and Mars were used to focus the telescope, and pointing errors were corrected to an accuracy of $\sim$3'' by performing frequent measurements of nearby pointing calibrators (namely quasars such as QSO B1730-130 and QSO J0854+2006, but also Neptune, the ionized region K3-50A, and the planetary nebula NGC 7027). The subreflector was wobbled with a throw of 120$^{\prime \prime}$ every 2 seconds to provide stable and flat baselines. The data were calibrated in units of main-beam temperature, T$_\mathrm{mb}$ by frequent (15-20 min) calibration scans using the chopper wheel method. The T$_{mb}$ values so obtained were then re-scaled using sources with stable, well-known fluxes (CW Leo and NGC 7027). The estimated flux calibration accuracy is about 20\%.

Integration times varied from an hour for $^{13}$CO lines to 6 hours for HCN, HCO$^+$, and HNC. Representative integration times are listed in Table \ref{Positions}. Weak lines were observed for significantly longer durations, up to 33 hours. System temperatures ranged between 102.4~K and 445.9~K, depending on the band and weather conditions.

\begin{table}
	\begin{center}
	\caption{Molecular Transitions Detected}
	\label{Transitions}
	\begin{tabular}{l c r}
	\hline
	\hline
	Molecule & Transition & $\nu$ (GHz) \\ 
	\hline
	CO 		&	$J=$ 1 $\rightarrow$ 0 &	115.2712018\phantom{$^*$} \\ %(2018) \\
			&	$J=$ 2 $\rightarrow$ 1 &	230.5380000\phantom{$^*$} \\ %(0000) \\
	$^{13}$CO &	$J=$ 1 $\rightarrow$ 0 &	110.2013543$^*$ \\ %(3543) \\
			&	$J=$ 2 $\rightarrow$ 1 &	220.3986842$^*$ \\ %(6842) \\
	C$^{17}$O &	$J=$ 1 $\rightarrow$ 0 &	112.3587770$^*$ \\ %(7770) \\
			&	$J=$ 2 $\rightarrow$ 1 &	224.7141870$^*$ \\ %(1870) \\
	HCN		&	$J=$ 1 $\rightarrow$ 0 &	88.6316022$^*$ \\ %(8470) \\
	HNC		&	$J=$ 1 $\rightarrow$ 0 &	90.6635680\phantom{$^*$} \\ %(5680) \\
	CN		&	$N=$ 1 $\rightarrow$ 0 &	113.1441573\phantom{$^*$} \\
			&					    &	113.1704915\phantom{$^*$} \\ %(4915) \\
			&					    &	113.1912787\phantom{$^*$} \\
			&					    &	113.4909702\phantom{$^*$} \\ %(9702)* \\
			&	$N=$ 2 $\rightarrow$ 1 &	226.6595584$^*$\\ %(5580) \\
			&					    &	226.8747813$^*$\\ %(1910) \\
	HCO$^+$	&	$J=$ 1 $\rightarrow$ 0 &	89.1885247\phantom{$^*$} \\ %(5247) \\
	N$_2$H$^+$ & $J=$ 1 $\rightarrow$ 0 &	93.1747000$^*$ \\ %(7000) \\
%									235.3800460 NOT FOUND
	CO$^+$	&	$J=$ 2 $\rightarrow$ 1 &	235.7896410\phantom{$^*$} \\ %(5530) \\
			&					   &	236.0625530\phantom{$^*$} \\ %(6410) \\
	\hline
	\end{tabular}
	\end{center}
	%\scriptsize{
	\tablefoot{\\
		Values of frequencies of hyperfine transitions obtained from the CDMS database. Line frequencies were measured in the laboratory by \cite{Muller01}, with updated catalog by \cite{Muller05}. \\
		Asterisks indicate frequencies representative of hyperfine transition complexes.
			}
\end{table}	

Scan averaging and baseline subtraction of measurements were performed with the line analysis software available in CLASS %\footnote{https://www.iram.fr/IRAMFR/GILDAS/doc/html/\allowbreak{class-html/class.html}} 
and through targeted inspection. Baseline oscillations were observed in some higher frequency spectra, but did not significantly affect measurements of individual molecular transitions.

%%%%%%%%%%%%%%% RESULTS %%%%%%%%%%%%%%%
\section{Results and Analysis}
\subsection{Molecular Line Measurements}

\longtab{
\begin{longtable}{l l c c c c c}
\caption{PNe Previously Surveyed in Molecular Lines} \\
\hline\hline
	PNe & Species & T$_{mb}$ ($\sigma_{mb}$) & $\int \, T_{mb} \, dV$ ($\sigma_{area}$) & Prev. Work  & Prev. Work  & Notes \\
		&		&		(K)		 	     & (K km s$^{-1}$)		    & T$_{mb}$ &  $\int \, T_{mb} \, dV$	& \\
	\hline
	NGC 7027 & CO (1-0)		& 9.6 (0.03) & 267.0 (0.14)	& 11.9 (---) & 332 (---)	& a \\
			& CO (2-1)		& 26.2 (0.05) & 520.0 (0.15)	& 30.9 (---) & 667 (---)	& a \\
			& $^{13}$CO (1-0)	& 0.28 (0.01) & 6.68 (0.06)	& 0.3 (---) & 6.4 (0.3)		& a, b \\
			& $^{13}$CO (2-1)	& 0.89 (0.01) & 22.4 (0.09)	& 1.3 (---) & 17.3 (0.8)	& a, b \\
			& C$^{17}$O (1-0)	& 0.045 (0.031)	& 1.51 (0.18) 	& --- (---)	& 0.92 (0.12)	& c \\
			& C$^{17}$O (2-1)	& 0.268 (0.081) & 3.33 (0.354) 	&  0.032 (---) & 1.87 (0.08)& d, c \\
			& CO$^+$ (2-1)		& 0.05 (0.01) & 2.18 (0.11) 	& 0.020 (---) & 0.87 (0.13) & e \\
			& N$_2$H$^+$ (1-0)	& 0.08 (0.01) & 0.640 (0.020)	& 0.017 (---) & 0.35 (---)	& d \\
			& HCO$^+$ (1-0)	& 1.36 (0.009) & 28.1 (0.05) 	& 0.247 (---) & 26.5 (0.2) 	& d, b \\
			& CN (2-1)		& 1.47 (0.016) & 86.5 (0.12) 	& 0.028 (---) & 41.9 (0.6) 	& d, b \\
			& CN (1-0)		& 0.543 (0.038) & 6.95 (0.048)	& --- (---)	& 25.1 (0.4)	& b \\
			& HCN (1-0)		& 0.35 (0.005) & 11.6 (0.03) 	& 0.059 (---) & 14.3 (0.4) 	& d, b \\
			& HNC (1-0)		& 0.034 (0.006) & 0.309 (0.022)& --- (---) & 1.0 (1.2) 		& b \\
			\hline
	NGC 6720 & CO (1-0)		& --- (---) & --- (---) 			& 0.047 (0.005) & --- (---) 	& f \\
			& CO (2-1)		& --- (---) & --- (---) 			& 0.023 (0.004) & 20.0 (0.4) & f, b\\
			& $^{13}$CO (1-0)	& --- (---) & --- (---) 			& --- (---) & 0.27 (0.11) 	& b \\
			& $^{13}$CO (2-1)	& 0.156 (0.008) & 1.30 (0.03) 	& --- (---) & 0.9 (0.3) 		& b \\
			& CO$^+$ (2-1)		& $<$0.021 (0.007) & $<$0.11 (0.04) & --- (---) & --- (---) 	& \\
			& N$_2$H$^+$ (1-0)	& $<$0.014 (0.006) & $<$0.012 (0.004) &	 --- (---) & --- (---)& \\
			& HCO$^+$ (1-0)	& 0.103 (0.004) & 1.760 (0.019) & 0.019 (0.002) & 0.86 (0.11) & f, b \\
			& CN (2-1)		& 0.31 (0.01) & 10.10 (0.06)	 & --- (---) & 4.5 (0.7) 	& b \\
			& CN (1-0)		& --- (---) & --- (---) 			& --- (---)	& 3.9 (0.3)		& b \\
			& HCN (1-0)		& 0.07 (0.004) & 1.83 (0.02) 	& --- (---) & 3.2 (0.2) 		& b \\
			& HNC (1-0)		& 0.04 (0.004) & 0.67 (0.02) 	& --- (---) & 0.74 (0.10) 	& b \\
			\hline
	NGC 6720 Rim & CO (1-0)	& --- (---) & --- (---) 			& 0.047 (0.005) & --- (---) 	& f \\
			& CO (2-1)		& --- (---) & --- (---) 			& 0.023 (0.004) & 20.0 (0.4) & f, b \\
			& $^{13}$CO (1-0)	& --- (---) & --- (---) 			& --- (---) & 0.27 (0.11) 	& b \\
			& $^{13}$CO (2-1)	& 0.05 (0.01) & 0.57 (0.06) 	& --- (---) & 0.9 (0.3) 		& b \\
			& CO$^+$ (2-1)		& $<$0.027 (0.009) & $<$0.18 (0.06) & --- (---)  & --- (---) 	& \\
			& N$_2$H$^+$ (1-0)	& $<$0.014 (0.006) & $<$0.017 (0.04) & --- (---) & --- (---)	& \\
			& HCO$^+$ (1-0)	& 0.063 (0.005) & 1.28 (0.03) 	& 0.019 (0.002) & 0.86 (0.11) &f, b \\
			& CN (2-1)		& 0.29 (0.01) & 11.1 (0.08) 	& --- (---) & 4.5 (0.7) 		& b \\
			& CN (1-0)		& --- (---) & --- (---) 			& --- (---)	& 3.9 (0.3)		& b \\
			& HCN (1-0)		& 0.074 (0.005) & 2.08 (0.03) 	& --- (---)  & 3.2 (0.2) 	& b\\
			& HNC (1-0)		& 0.027 (0.005) & 0.603 (0.027) & --- (---) & 0.74 (0.10) 	& b \\
			\hline
	NGC 7293 & CO (1-0)		& 0.66 (0.02) & 6.0 (0.06) 		& 0.17 (0.030) & --- (---) 	& g \\
			& CO (2-1)		& 2.05 (0.06) & 12.3 (0.1) 		& 0.40 (0.050) & 17.5 (0.7) & g, b \\
			& $^{13}$CO (1-0)	& 0.044 (0.009) & 0.096* (0.014) & --- (---) & 0.67 (0.09) 	& b \\
			& $^{13}$CO (2-1)	& 0.047 (0.006) & 0.60 (0.03) 	& --- (---) & 1.8 (0.2)  	& b \\
			& CO$^+$ (2-1)		& $<$0.021 (0.007) & $<$0.07 (0.02) & --- (---) & --- (---) 	& \\
			& N$_2$H$^+$ (1-0)	& $<$0.001 (0.004) & --- (---)	& --- (---) & --- (---)		& \\
			& HCO$^+$ (1-0)	& 0.088 (0.044) & 1.020 (0.017) & 0.04 (0.008) & 0.89 (0.10) & g, b  \\
			& CN (2-1)		&  0.029 (0.009) & 0.18 (0.02) 	& --- (---) & 1.0 (0.2) 		& b \\
			& CN (1-0)		& 0.054 (0.008) & 0.629 (0.04) 	& --- (---)	& 2.7 (0.2)		& b \\
			& HCN (1-0)		& 0.041 (0.003) & 0.649 (0.014) & --- (---) & 1.6 (0.2) 	& b \\
			& HNC (1-0)		& 0.038 (0.003) & 0.465 (0.013) & --- (---) & 0.80 (0.11) 	& b \\
			\hline
	NGC 6781 & CO (1-0)		& 1.14 (0.02) & 15.3 (0.07) 	& 0.35 (---)  & --- (---) 	& h \\
			& CO (2-1)		& 5.26 (0.05) & 49.6 (0.11) 	& --- (---) & 28.4 (0.6) 	& b \\
			& $^{13}$CO (1-0)	& 0.049 (0.009) & 0.159 (0.016) & --- (---) & 0.32 (0.14) 	& b \\
			& $^{13}$CO (2-1)	& 0.203 (0.007) & 2.21 (0.03) 	& --- (---) & 1.7 (0.2) 		& b \\
			& CO$^+$ (2-1)		& $<$0.021 (0.007) & $<$0.09 (0.03) & 0.031 (0.008) & 0.249 (---) & i \\
			& N$_2$H$^+$ (1-0)	& $<	$0.011 (0.003) &--- (---)	& --- (---) & --- (---)		& \\
			& HCO$^+$ (1-0)	& 0.167 (0.003) & 3.520 (0.018) & --- (---) & 2.2 (0.2) 	& b \\
			& CN (2-1)		& 0.40 (0.01) & 13.2 (0.06) 	& --- (---) & 4.6 (0.5) 		& b \\
			& CN (1-0)		& 0.317 (0.042) & 5.51 (0.33) &	 --- (---)	& 7.0 (0.4)		& b \\
			& HCN (1-0)		& 0.16 (0.003) & 3.83 (0.02) 	& --- (---) & 2.4 (0.2) 		& b \\
			& HNC (1-0)		& 0.08 (0.003) & 1.73 (0.02) 	& --- (---) & 1.65 (0.15) 	& b \\
	\hline
	\label{Tpeaks1}
\end{longtable}
%	\tablefoot{\\
			*: Uncertain\\
			a: \citet{SG12}; 
			b: \citet{Bachiller97}; 
			c: \citet{Kahane92}; 
			d: \citet{Zhang08}; 
			e: \citet{HK01}; 
			f: \citet{Edwards14}; 
			g: \citet{Zack13}; 
			h: \citet{Zuckerman90}; 
			i: \citet{Bell07}.
%			}
}
%%%%% Mostly Unobserved %%%%%
\longtab{
\begin{longtable}{l l c c | c c c}
\caption{PNe Not Previously Surveyed in Molecular Lines} \\
\hline\hline
	PNe & Species & T$_{mb}$($\sigma_{mb}$) & $\int \, T_{mb} \, dV$ ($\sigma_{area}$) & Prev. Work  & Prev. Work  & Notes \\
		&		&		(K)		 	     & (K km s$^{-1}$)		    & T$_{mb}$ &  $\int \, T_{mb} \, dV$	& \\
	\hline
	BD+30$\degree$3639 	& CO (1-0) & $<$0.39 (0.13) 		& $<$2.5 (0.8) & --- (---) & 5.2 (---) & a \\
			& CO (2-1)	& $<$1.2 (0.4) & $<$5.6 (1.8) 		& $\leq$0.20 (---) & 4.7 (---) & b, a \\
			& $^{13}$CO (1-0)	& $<$0.09 (0.03) & $<$0.57 (0.19) & --- (---) & --- (---)  	& \\
			& $^{13}$CO (2-1)	& $<$0.018 (0.006) & $<$0.17 (0.06) & --- (---) & --- (---) 	&  \\
			& CO$^+$ (2-1)	& $<$0.018 (0.006) & $<$0.16 (0.05) & --- (---) & --- (---) 		& \\
			& HCO$^+$ (1-0)& 0.029 (0.003) & 1.16 (0.02) 		& --- (---) & --- (---) 		& \\
			& CN (2-1)	& $<$0.027 (0.009) & $<$0.17 (0.06) & --- (---) & --- (---) 		& \\
			& CN (1-0)	& $<$0.1 (0.05) & --- (---) 			& --- (---) & --- (---) 		&  \\
			& HCN (1-0)	& $<$0.009 (0.003) & $<$0.07 (0.02) & --- (---) & --- (---) 		& \\
			& HNC (1-0)	& $<$0.009 (0.003) & $<$0.06 (0.02) & --- (---) & --- (---) 		& \\
			\hline
	NGC 6445 & CO (1-0)	& 0.23 (0.03) & 6.36 (0.14) 		& 0.25 (---) & 9.29 (---) 	& c \\
			& CO (2-1)	& 1.24 (0.07) & 14.8 (0.17) 		& 0.25 (---) & 11.1 (---) 	& d, c \\
			& $^{13}$CO (1-0)	& 0.03 (0.01) & 0.051* (0.014) 	& --- (---) & --- (---) 		& \\
			& $^{13}$CO (2-1)	& 0.32 (0.006) & 0.93 (0.05) 	& --- (---) & --- (---) 		& \\
			& CO$^+$ (2-1)	& $<$0.021 (0.007) & $<$0.1 (0.03) 	& --- (---) & --- (---) 		& \\
			& N$_2$H$^+$ (1-0)	& 0.006 (0.002) & 0.025 (0.007) & --- (---) & --- (---)		& \\
			& HCO$^+$ (1-0)& 0.052 (0.003) & 1.78 (0.02) 		& --- (---) & --- (---) 		& \\
			& CN (2-1)	&  0.257 (0.009) & 11.2 (0.06) 		& --- (---) & --- (---) 		& \\
			& CN (1-0)	& 0.362 (0.044) & 2.74 (0.21)		& --- (---) & --- (---)		&  \\
			& HCN (1-0)	& 0.115 (0.003)  & 4.54 (0.02) 		& --- (---) & --- (---) 		& \\
			& HNC (1-0)	&  0.034 (0.003) & 1.12 (0.02) 		& --- (---) & --- (---) 		& \\
			\hline
	NGC 7008 & CO (1-0)	& 1.15 (0.03) & 2.56 (0.04) 		& --- (---) & --- (---) 		& \\
			& CO (2-1)	& 1.64 (0.06) & 2.09* (0.05) 		& --- (---) & --- (---)  		& \\
			& $^{13}$CO (1-0)	& 0.057 (0.009) & 0.115* (0.013) & --- (---) & --- (---) 		& \\
			& $^{13}$CO (2-1)	& 0.04* (0.02) & 0.55* (0.09) 	& --- (---) & --- (---) 		& \\
			& CO$^+$ (2-1)	& $<$0.06 (0.02) & $<$0.08 (0.03) 	& --- (---) & --- (---) 		& \\
			& N$_2$H$^+$ (1-0)	& $<$0.01 (0.006) & --- (---) 	& --- (---) & --- (---) 		& \\
			& HCO$^+$ (1-0)& 0.013 (0.005) & 0.113 (0.015) 	& $<$0.004 (---) & --- (---) & e \\
			& CN (2-1)	& $<$0.06 (0.02) & $<$0.07 (0.02) 	& --- (---) & --- (---) 		& \\
			& CN (1-0)	& $<$0.016 (0.008) & $<$0.082 (0.025) & --- (---)	& --- (---)	&  \\
			& HCN (1-0)	& $<$0.015 (0.005) & $<$0.019 (0.006) & $<$0.006 (---) & --- (---) & e \\
			& HNC (1-0)	& $<$0.015 (0.005) & $<$0.019 (0.006) & --- (---) & --- (---) 	& \\
			\hline
	NGC 6853 & CO (1-0)	& 0.13 (0.02) & 1.41 (0.07) 		& 0.097 (0.012) & 1.21 (0.29) & f \\
			& CO (2-1)	& 0.36( 0.06) & 3.48* (0.14) 		& 0.251 (0.008) & 2.49 (0.06) & f \\
			& $^{13}$CO (1-0)	& $<$0.03 (0.01) & $<$0.09 (0.03) & --- (---) & --- (---) 	& \\
			& $^{13}$CO (2-1)	& 0.013 (0.004) & 0.09 (0.02) 	& --- (---) & --- (---) 		& \\
			& CO$^+$ (2-1)	& $<$0.012 (0.004) & $<$0.052 (0.018) & --- (---) & --- (---)  	& \\
			& N$_2$H$^+$ (1-0)	& $<$0.005 (0.002) & $<$0.006 (0.002) & --- (---) & --- (---) & \\
			& HCO$^+$ (1-0)	& 0.020 (0.002) & 0.096 (0.006) 	& 0.026 (0.002) & 0.33 (0.03) & f \\
			& CN (2-1)	& 0.072 (0.007) & 1.38* (0.1) 		& --- (---) & --- (---) 		& \\
			& CN (1-0)	& 0.108 (0.027) & 0.524 (0.053)	& --- (---)	& --- (---)		& \\
			& HCN (1-0)	& 0.069 (0.002) & 0.637 (0.008) 	& --- (---) & --- (---) 		& \\
			& HNC (1-0)	& 0.030 (0.002) & 0.204 (0.007) 	& --- (---) & --- (---) 		& \\
			\hline
	NGC 6772 & CO (1-0)	& 0.24 (0.02) & 2.5 (0.05) 			& 0.03 (---) & --- (---) 		& g \\
			& CO (2-1)	& 0.55 (0.04) & 6.36 (0.11) 		& 0.05 (0.01) & 1.39 (0.41*) & e \\
			& $^{13}$CO (1-0)	& $<$0.021 (0.007) & $<$0.07 (0.02) & --- (---) & --- (---) 	& \\
			& $^{13}$CO (2-1)	& 0.040 (0.007) & 0.32 (0.03) 	& --- (---) & --- (---)  		& \\
			& CO$^+$ (2-1)	& $<$0.021 (0.007) & $<$0.1 (0.03) 	& --- (---) & --- (---)  		& \\
			& N$_2$H$^+$ (1-0)	& $<$0.01 (0.004) &$<$0.005 (0.002) & --- (---) & --- (---) 	& \\
			& HCO$^+$ (1-0)	& 0.030 (0.003) & 0.358 (0.012) 	& $<$0.003 (---) & --- (---) 	& e \\
			& CN (2-1)	& 5.89 (0.01) & 1.26 (0.005) 		& --- (---) & --- (---) 		& \\
			& CN (1-0)	& 0.040 (0.010) & 0.329 (0.037)	& --- (---) & --- (---) 		&  \\
			& HCN (1-0)	& 0.021 (0.003) & 0.419 (0.016) 	& 0.010 (0.006) & 0.299 (0.247*) & e \\
			& HNC (1-0)	& 0.012 (0.003) & 0.153 (0.013) 	& 0.0027 (0.0012) & --- (---) 	& h  \\
			\hline
	\label{Tpeaks2}
\end{longtable}
%	\tablefoot{\\
			*: Uncertain\\
			a: \citet{Bachiller91}; % (IRAM 30m)\\
			b: \citet{Gussie95};  %(JCMT) \\
			c: \citet{Sun00};  % (Purple Mountain Observatory)\\
			d: \citet{Huggins89};  % (NRAO 12m)\\
			e: \citet{Schmidt16};  %(ARO 12m, ARO 10m SMT)\\
			f: \citet{Edwards14}; 
			g: \citet{Zuckerman90};  %(NRAO 12m)\\
			h: \cite{Schmidt17}. %(ARO 12m)
%			}
}

Observations of the thirteen molecular lines detectable in the 88-236 GHz range provide new data for the nine PNe in our survey, including the well observed NGC 7027.
We identified and compiled a list of these molecular transitions, with lines of primary interest and their measured intensities summarized in Tables \ref{Tpeaks1} and \ref{Tpeaks2}, respectively, for objects that have been well observed in the past and those for which this survey yielded a significant number of new detections.
%Notations of lines that have been previously detected, as well as the relevant papers, are also included.
The tables list peak flux (T$_{mb}$) as defined by the main beam temperature, integrated line intensity ($\int \, T_{mb} \, dV$), reported parameters for lines detected in past studies, and the (1 $\sigma$) formal errors for each measurement. The values of temperature, intensity, and their errors were obtained from the Gaussian fitting procedure in CLASS.

\subsection{Survey Overview}

We present the first detections of $^{13}$CO (1-0, 2-1), HCO$^+$, CN, HCN, and HNC in NGC 6445 (Section \ref{6445}) as well as HCO$^+$ in BD+30$\degree$3639 (Section \ref{BD+30}). Transitions of $^{13}$CO (2-1), CN, HCN, and HNC were also detected for the first time in NGC 6853 (Section \ref{6853}) as well as $^{13}$CO (2-1), CN, and HCO$^+$ in NGC 6772 (Section \ref{endPN}). The individual PNe discussed in Sections \ref{beginPN}-\ref{6781} have been the subject of previous radio molecular line surveys \citep{Bachiller97, Edwards14, Zhang08}, and the present survey provided only a few new detections for these objects. 
Among the molecular ions surveyed, HCO$^+$ was detected in most of the PNe, including the aforementioned new detections in three objects. However, the only PN unambiguously detected in CO$^+$ and N$_2$H$^+$ was NGC 7027 (Section \ref{beginPN}), which was previously known to exhibit emission from both molecules \citep{HK01, Zhang08}.

Figures \ref{stacked7027} through \ref{stacked6772} illustrate the emission lines detected in each objects and compare their relative strengths. Across PNe with detected CN, intensities of the strongest line scale more or less consistently with lesser CN hyperfine lines. With this in mind, we only list parameters for the brightest CN (1-0) hyperfine components in Tables \ref{Tpeaks1} and \ref{Tpeaks2}. Hydrogen recombination lines were also detected in two PNe, NGC 7027 and BD+30$\degree$3639, and these lines are displayed in Figure \ref{Hydrogen}. Both objects are very young and compact, with high ionized gas densities capable of producing strong atomic hydrogen emission. All CO isotopologue lines for NGC 7008 (Section \ref{7008}) were also previously unobserved, however the detections obtained here are likely due to ISM gas along the line of sight to the PN.

We find line strengths for the PNe in our survey compare favorably with previous observations \citep[e.g.,][]{Bachiller97}. Because many of the objects were observed at different positions, we limit direct comparisons to a pair of representative PNe. In NGC 7027, we find line ratios of $^{12}$CO/$^{13}$CO and HCO$^{+}$/$^{13}$CO that are within 8\% of those reported by \cite{Bachiller97}, while CN/$^{13}$CO is within 23\%. Similarly, we noted a difference of only 15\% in the reported $^{12}$CO/$^{13}$CO line ratio in NGC 6781, despite an offset of 50$^{\prime \prime}$ between the positions observed within this PN in the two surveys.

\subsection{Individual Planetary Nebulae}
\subsubsection{NGC 7027} %%%%% NGC 7027 %%%%%
\label{beginPN}

	%Figures of 7027 stacked molecular lines
	\begin{figure}
		%\centering
		\resizebox{\hsize}{!}{\includegraphics[trim={8cm 2.5cm 8cm 2cm}, clip, width=0.7\textwidth]{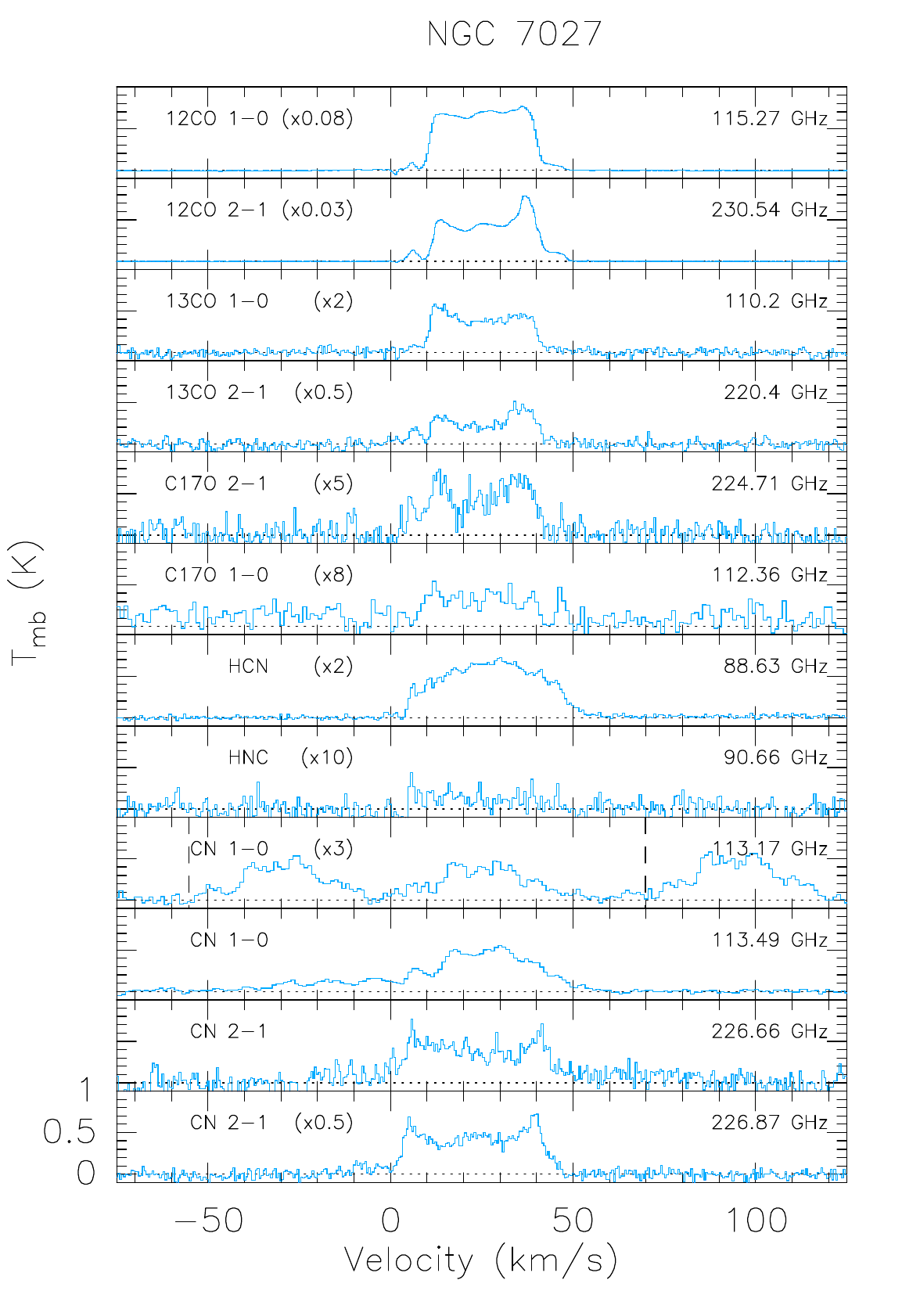}}
				 %trim={<left> <lower> <right> <upper>}
		\caption{Spectra for transitions of neutral molecules detected in the survey data for NGC 7027. Individual spectra have been scaled to facilitate comparison, and vertical dashed lines have been added to indicate the systemic velocity positions for the hyperfine CN lines. 
		The x-axis indicates velocity with respect to the local standard of rest (v$_{LSR}$) and the y-axis is main-beam antenna temperature (K).}
		\label{stacked7027}
	\end{figure}

	\begin{figure}
		\centering
		\resizebox{\hsize}{!}{\includegraphics[trim={3cm 5cm 3cm 3cm}, clip, width=0.7\textwidth]{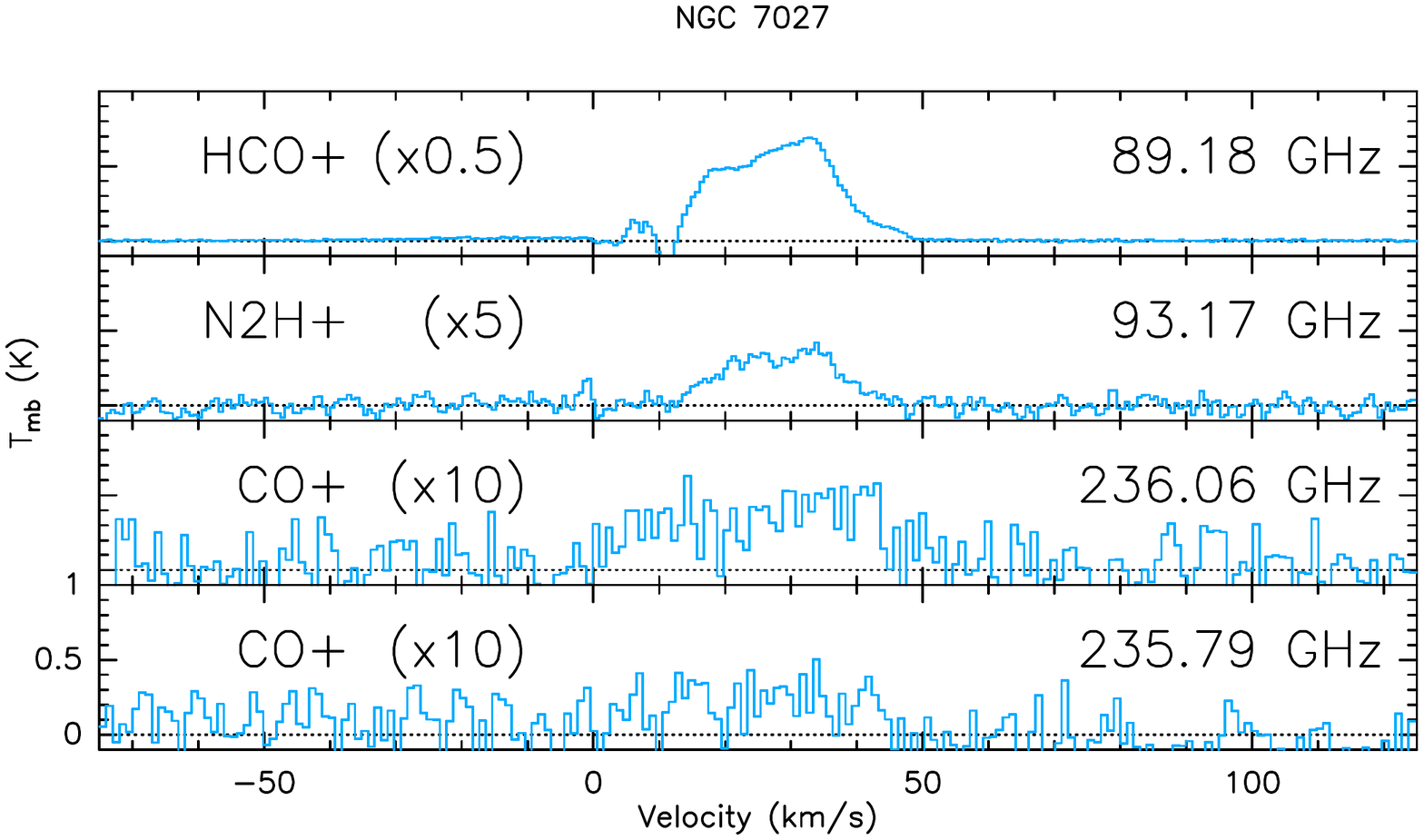}}
		\caption{As in Figure \ref{stacked7027}, but for transitions of molecular ions detected in the survey data for NGC 7027.}
		\label{stacked7027_ion}
	\end{figure}

\indent The data obtained for NGC 7027 yield by far the richest molecular line spectrum among the objects included in this survey. This young, bright nebula has a notably hot CSPN \citep[$T_\star$=175,000 K,][]{Kastner12, Frew08} and displays an abundance of atomic and molecular emission lines \citep{HK01}. With a dynamical age of only 700 years \citep{Latter00} and a high CSPN mass (0.7 M$_\odot$), NGC 7027 is a rapidly evolving nebula \citep{Zhang08}. The morphology of the H{\sc ii} region is that of a prolate ellipsoidal PN, with a distinct waist and symmetrical outflow lobes visible in the infrared and X-ray \citep{Graham93, Kastner01}. The clover-like symmetrical shell of molecular hydrogen forms a wispy and filamentary structure about the elongated shell of ionized gas \citep{Latter00, Cox02}. The H$_2$ emission further traces out the thin PDR, separating the inner ionized gas from the expanding molecular envelope, which has been modeled spatio-kinematically by \cite{SG12} and determined to have a total molecular mass of roughly 1.3 $M_\odot$. The presence of point-symmetric holes in the H$_2$ region also indicates the presence of collimated outflows, a common phenomenon in young PNe \citep{Cox02}.

%12CO (1-0), 12CO (2-1), 13CO (1-0), 13CO (2-1), C17O(2-1), C17O(1-0), HCN, HNC, CN(2-1), CN(1-0), HCO+, CO+, N2H+
All thirteen molecular transitions surveyed have been detected in NGC 7027 and their individual spectra are displayed in Figures \ref{stacked7027} and \ref{stacked7027_ion}, which illustrate detections of neutral and ionized molecules, respectively. From the shapes of the spectral emission lines, several characteristics of this PN can be discerned. The distinct double-peaked structure found in CO isotopologues and 1~mm lines traces the molecular gas with high line-of-sight velocities. Because the molecular shell of NGC 7027 extends to >70$^{\prime \prime}$ \citep{HK01}, the $\sim$10$^{\prime \prime}$ beam of the 30 m at 1~mm misses much of the outer envelope gas emitting near the systemic velocity. In contrast, the larger beam size of our 3~mm observations more completely samples the entire shell of the nebula, resulting in a larger contribution from material expanding along the plane of the sky and, hence, flatter-topped line profiles. The same explanation applies to the difference between the line profiles measured here and those measured by \cite{Zhang08} with the larger beam of the ARO 12 m telescope. Indeed, our CO isotopologue spectral line shapes are consistent with 30 m observations by \cite{Herpin02}.

The central velocity of NGC 7027 was estimated by averaging the widths of each profile and assuming symmetric expansion. The CO spectral line profiles show blueshifted and redshifted edges at 10.7 and 36.9 km s$^{-1}$, respectively. We thereby obtain a systemic velocity of  23.8 km s$^{-1}$ and an expansion velocity of 13.1 km s$^{-1}$. These results are consistent with those of \cite{Graham93}, \cite{Cox02}, and \cite{Herpin02}, all of whom determined the systemic velocity of NGC 7027 to be approximately 25 km s$^{-1}$ and found molecular outflow velocities of 15 km s$^{-1}$. The nitrogen-bearing molecules, however, present a slightly faster expanding shell, with blueshifted and redshifted outflow velocities of 5 and 40 km s$^{-1}$, respectively. 
The velocity difference may be attributed to the N-bearing molecules being more abundant in the collimated, higher-velocity outflows present in NGC 7027.

\begin{figure}
	\centering
	\resizebox{\hsize}{!}{\includegraphics[trim={2cm 2.5cm 3cm 2cm}, clip, width=\textwidth]{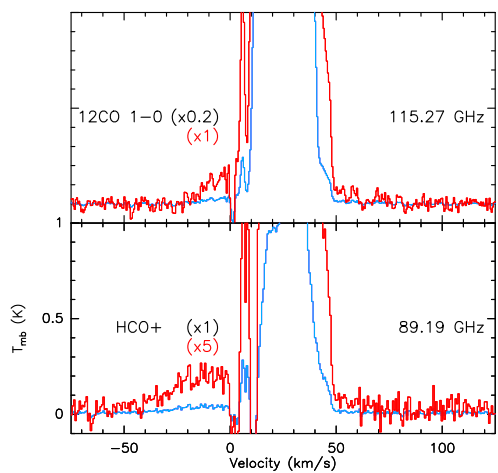}}
	\caption{Expanded spectra of CO and HCO$^+$ in NGC 7027, illustrating the presence of multiple absorption features and an extended blue wing.}
	\label{7027_Wing}
\end{figure}

We also observe evidence of self-absorption in the HCO$^+$ spectrum and possibly in CO (1-0); on the blueshifted side of the HCO$^+$ emission line, there appear absorption features at $\sim$0.8, 3.2, and 11.0 km s$^{-1}$ that dip below the averaged continuum, which has been baseline-subtracted (Figure \ref{7027_Wing}). We believe these are circumstellar features and not interstellar as there is no observed extended emission in wide CO maps or self-absorption in higher CO J lines \citep{SG12}. 
Beyond the absorption features, an extended blueshifted wing is apparent within the HCO$^+$ and CO (1-0) spectra (Figure \ref{7027_Wing}). Extended $^{12}$CO wings were first detected in NGC 7027 by \cite{Bujarrabal01}. While the CO wing is less extensive, the HCO$^+$ wing extends to $\sim$-40 km s$^{-1}$, indicating the presence of a molecular outflow with velocity $\sim$65 km s$^{-1}$, i.e.,  nearly 5 times that of the overall molecular expansion velocity. This velocity is strictly a lower limit, dependent on the angle of observation. This suggests that low excitation, approaching regions of the outer molecular envelope are absorbing the line and continuum emission from molecules of high abundance in the inner regions of the PN.

We confirm the detection of the ion N$_2$H$^+$ made by \cite{Zhang08}. Formation of N$_2$H$^+$ could be a result of the strong presence of X-rays from the CSPN, or due to shocks in the pPN phase.

While the double-peaked structure that is characteristic of an expanding envelope of molecular gas can be seen in most of the emission lines observed, it is notably absent in HCN, due to its hyperfine splitting at closely spaced frequencies (88.630, 88.631, and 88.634 GHz). Line broadening from expansion blends these hyperfine features. Close hyperfine lines may also be blended and thus unresolved in $^{13}$CO, C$^{17}$O, and N$_2$H$^+$. CN displays hyperfine structure as well (Table \ref{Transitions}), but at generally separable frequencies for both the N=2-1 and N=1-0 transitions. Two of the three hyperfine components of CO$^+$ are detected (Figure \ref{stacked7027_ion}).

\begin{figure}
		\centering
		\resizebox{\hsize}{!}{\includegraphics[trim={0cm 0cm 8cm 0cm}, clip, width=0.7\textwidth]{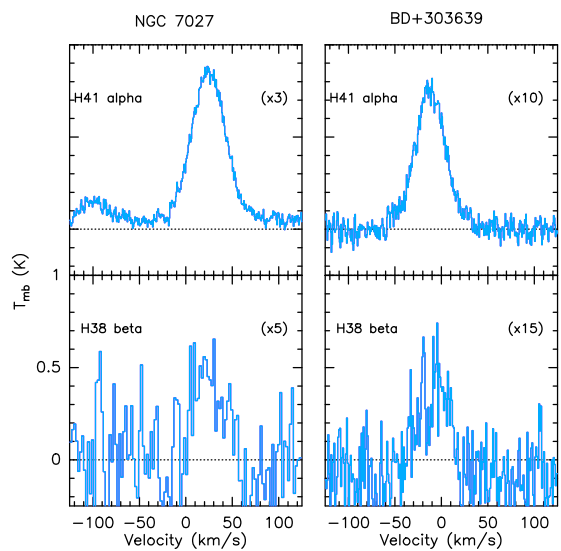}}
		\caption{Hydrogen recombination lines detected in two of the surveyed PNe, NGC 7027 and BD+303639.} %High energy level lines of H41$\alpha$ and H38$\beta$ indicate strong ionization from the central stars of these objects. This trend is furthered by the fact that both PNe are young and their ionized regions are fully contained within beams centered on their CSPN.}
		\label{Hydrogen}
	\end{figure}

Further analysis of the spectra revealed detections of hydrogen recombination lines at 92.0 (H41$\alpha$) and 222 (H38$\beta$) GHz (Figure \ref{Hydrogen}). H41$\alpha$ has a measured peak intensity of 0.278 K and integrated flux of 4.12 K km s$^{-1}$, and was previously identified by \cite{Zhang08}. H38$\beta$ has a peak intensity of 0.096 K and an integrated flux of 2.63 K km s$^{-1}$. 

\subsubsection{NGC 6720} \label{Sect6720} %%%%% NGC 6720 and 6720 Rim %%%%%

	%Figure of 6720 stacked molecular lines
	\begin{figure}
		\centering
		\resizebox{\hsize}{!}{\includegraphics[trim={2.5cm 2.5cm 3cm 2cm}, clip, width=0.7\textwidth]{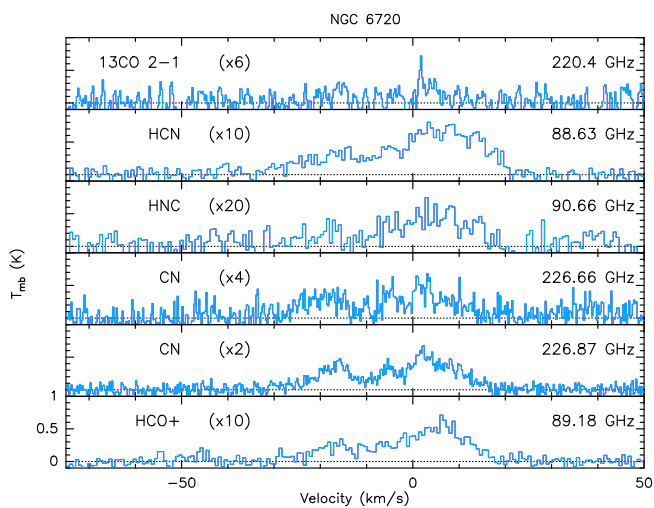}}
		\caption{As in Figure \ref{stacked7027}, molecular transitions detected in the survey data obtained toward the central star of NGC 6720.}
		\label{stacked6720}
	\end{figure}
	%Figure of 6720Rim stacked molecular lines
	\begin{figure}
		\centering
		\resizebox{\hsize}{!}{\includegraphics[trim={3cm 2.5cm 3cm 2cm}, clip, width=0.7\textwidth]{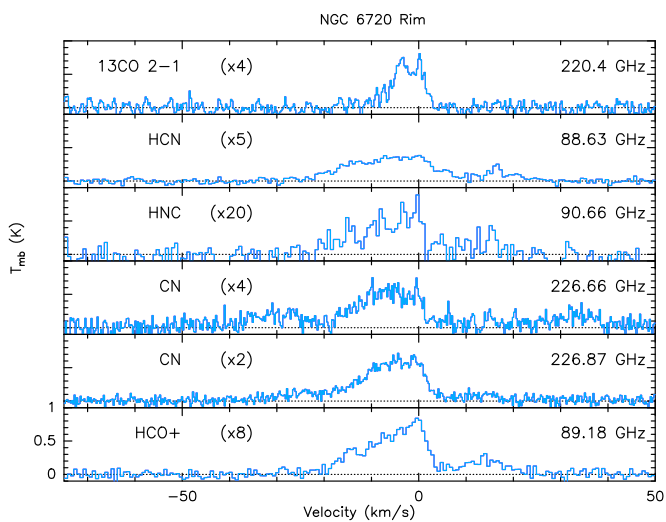}}
		\caption{As in Figure \ref{stacked7027}, molecular transitions detected in the survey data obtained toward the rim of NGC 6720, referred to throughout this paper as NGC 6720 Rim.}
		\label{stacked6720Rim}
	\end{figure}

\indent Commonly known as the Ring Nebula, NGC 6720 (M57) has been well studied in molecular gas \citep{Bachiller97, Edwards14}. Its dynamical age is roughly 7,000 years \citep{ODell07} and it does not have detectable X-ray emission \citep{Kastner12}.
We observed two positions within NGC 6720, one toward the CSPN, and one just outside the optically bright nebula (Figure \ref{PNePositions}). Emission lines for these regions are displayed in Figures \ref{stacked6720} and \ref{stacked6720Rim}, respectively. Given the extensive CO survey of NGC 6720 performed by \cite{Bachiller89}, we chose not to cover higher frequency wavebands (including three of the CO isotopologues) in our survey observations, and as such only six molecular transitions were detected.

The CSPN pointing (Figure \ref{stacked6720}) encompasses molecular gas associated with both the approaching and receding layers of the molecular shell, and as a result this pointing yields broad, double-peaked line profiles. We measure blueshifted and redshifted velocity components of -16.2 km s$^{-1}$ and 4.6 km s$^{-1}$, respectively, at the central star of NGC 6720. These values agree well with the CO measurements of \cite{Edwards14}, though they also detect a redshifted component at 15.7$\pm$2.6~km s$^{-1}$.

The spectra for the position offset from the CSPN yield line profiles that are dominated by emission near the systemic velocity of the PN, consistent with the expectation that the velocity of the bulk of the gas along the nebular rim is perpendicular to the line of sight. The extended tail toward the blueshifted side of the nebula indicates that the emission is dominated by the forward-directed side of the molecular region. At this rim position, emission peaks appear at -6.7 km s$^{-1}$ and -0.8 km s$^{-1}$. Another peak was identified at 15.3 km s$^{-1}$, but only in H-bearing molecules. This spectral feature in the Rim position corresponds to the receding emission component detected in molecular spectra obtained by \cite{Edwards14}. Although we did not observe CO (2-1) in NGC 6720, the observed profiles match the structure seen by \cite{Bachiller89} in their ($\Delta\alpha$=-40", $\Delta\gamma$=-20") position. Further, the average systemic velocity of NGC 6720 across both positions ($\sim$-0.8 km s$^{-1}$) agrees well with the measurements of \cite{Bachiller97}.% at -2 km s$^{-1}$.

\subsubsection{NGC 7293} %%%%% NGC 7293 %%%%%

	%Figure of 7293 stacked molecular lines
		\begin{figure}
			\centering
			\resizebox{\hsize}{!}{\includegraphics[trim={7cm 2.5cm 7cm 2cm}, clip, width=0.7\textwidth]{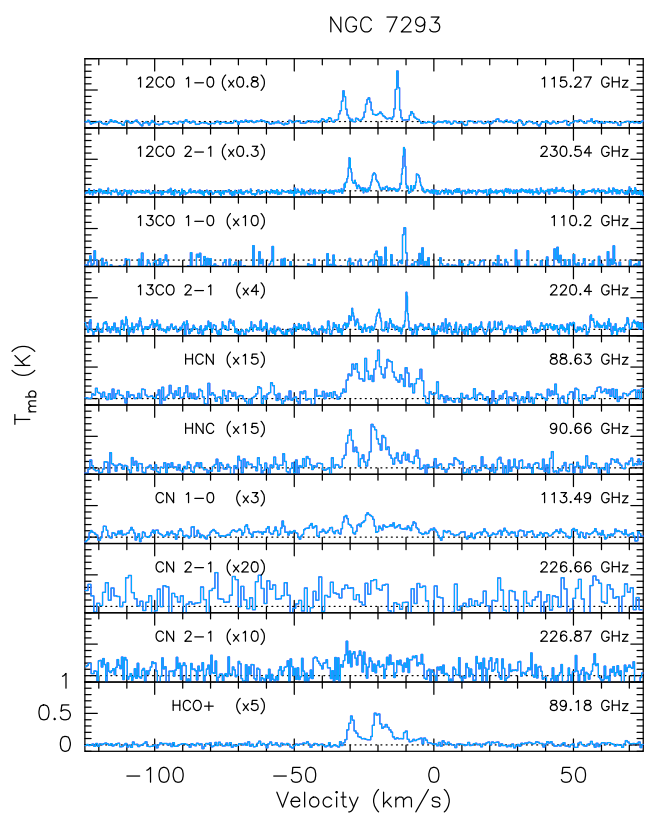}}
			\caption{As in Figure \ref{stacked7027}, molecular transitions detected in the survey data for NGC 7293.}
			\label{stacked7293}
		\end{figure}
		
\indent At a distance of only 0.20 kpc \citep{Gaia} and a radius of 0.46 pc, the Helix Nebula (NGC 7293) represents the largest and nearest PN in this study. It is a relatively evolved object with a dynamical age of 12,000 years \citep{Schmidt17}, and contains a hard, point-like X-ray source \citep{Guerrero01}. Early radio molecular line observations were carried out by \cite{Huggins92} and \cite{Bachiller97}, with many additional studies performed across the expansive molecular emission region of the PN \citep{Young99, Zack13, Zeigler13, Schmidt17, Schmidt18}.

Our observations of NGC 7293 were performed toward the edge of the molecular envelope (Figure \ref{PNePositions}). While all the molecular transitions studied in this paper have previously been detected in this PN, this work adds integrated flux values for these lines in a previously unobserved region of the object. As in previous single-dish mapping studies, there is significant structure present in the spectra (Figure \ref{stacked7293}). In particular, four distinct peaks appear in the $^{12}$CO (1-0, 2-1) transitions. This complex line profile can be attributed to multiple, distinct molecular knots along the line of sight sampled by the 30 m beam. Due to the large angular diameter of NGC 7293, the IRAM telescope beam measured only a small portion of the extensive molecular region. As such, it is only possible to make a comparison with a nearby region ($\Delta\alpha$=-435", $\Delta\gamma$=75"), as observed by \cite{Zack13}, \cite{Schmidt17}, and \cite{Schmidt18}. Their spectra revealed three CO velocity components at -11, -21, and -29 km s$^{-1}$ and -13, -21, -29 km s$^{-1}$ in HNC and HCN, respectively. This lines up well with our spectra, in which strong velocity components are observed at roughly -13, -23, and -32 km s$^{-1}$, with two other potential features at -8 and -19 km s$^{-1}$.
Observations by \cite{Schmidt18} of CN at the nearby position (-435, 75), as well as seven other positions across the Helix, detected hyperfine structure indicative of optically thin emission; our data are too noisy to confirm these results.

\subsubsection{NGC 6781} %%%%% NGC 6781 %%%%%
\label{6781}

	%Figure of 6781 stacked molecular lines
		\begin{figure}
			\centering
			\resizebox{\hsize}{!}{\includegraphics[trim={7cm 2.5cm 8cm 2cm}, clip, width=0.7\textwidth]{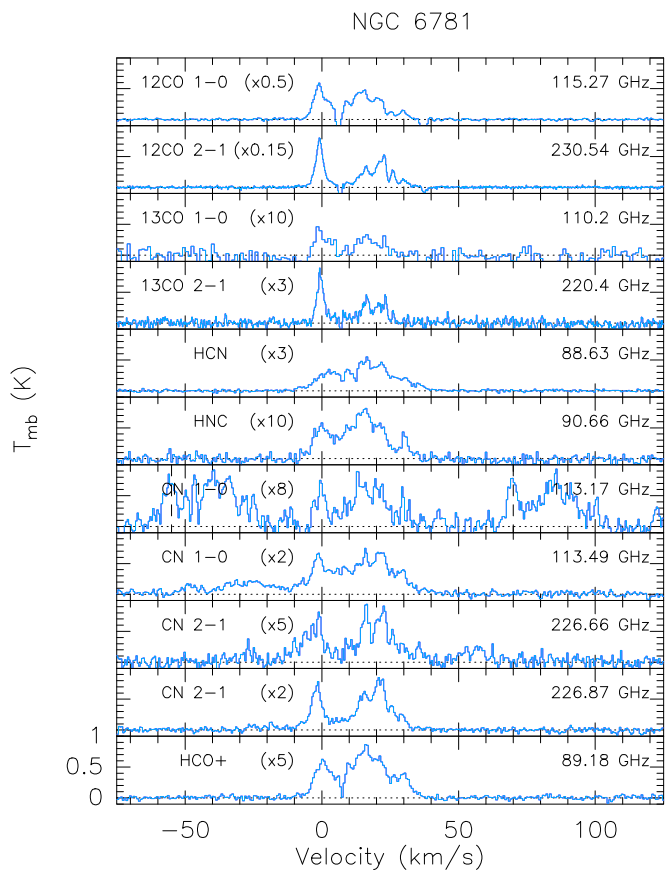}}
			\caption{As in Figure \ref{stacked7027}, molecular transitions detected in the survey data for NGC 6781. Vertical dashed lines indicate the systemic velocity positions for the hyperfine CN lines.}
			\label{stacked6781}
		\end{figure}

NGC 6781 is a highly evolved PN with an estimated dynamical age of at least 20,000 years \citep{Ueta14}, making it the oldest object included in our survey. As one of our three surveyed PNe not detected in X-rays, NGC 6781 acts as a control for any potential X-ray-induced chemistry effects.

Radio molecular line observations of NGC 6781 were previously carried out by \cite{Bachiller97}. New observations of the PN presented here probe a different region of the expanding shell. NGC 6781 was detected in all molecular transitions except CO$^+$ (Figure \ref{stacked6781}). The molecular line profiles of NGC 6781 display complex structures. At least four distinct velocity components were detected across the lines. Blueshifted and redshifted components at -0.3 and 29.4 km s$^{-1}$ respectively were detected for all molecular transitions. This suggests a systemic velocity of 14.6 km s$^{-1}$. A central component at 14.7 km s$^{-1}$ was also ubiquitously detected, with a fourth line at 21.4 km s$^{-1}$ identified in most of the transitions. These velocity profiles generally match up with those presented in \cite{Bachiller97}, who found velocity extremes of CO from -5 to 39 km s$^{-1}$.

The strong, narrow absorption component located at $\sim$6.5 km s$^{-1}$ in the CO line profiles is present across all detected transitions and is likely due to the presence of molecular cloud emission in the off-source (background sky) reference position.

\subsubsection{BD+30$\degree$3639} %%%%% BD+303639 %%%%%
\label{BD+30}

	%Figure of BD+30 stacked molecular lines
	\begin{figure}
		\centering
		\resizebox{\hsize}{!}{\includegraphics[trim={2cm 8cm 3cm 7cm}, clip, width=0.7\textwidth]{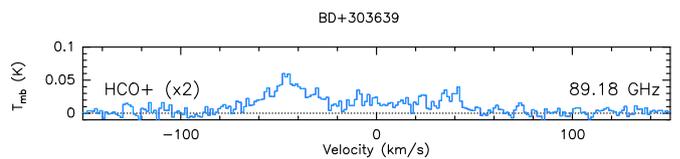}}
		\caption{As in Figure \ref{stacked7027}, the molecular transition detected in the survey data for BD+30$\degree$3639.}
		\label{stackedBD+}
	\end{figure}

\indent Also known as Campbell's Star, BD+30$\degree$3639 (hereafter BD+30), one of the youngest and most compact PN in our survey, has only just begun ionizing its envelope of ejected material within the past 1,000 years \citep{Li02}. The Wolf-Rayet-type central star (WC9) has the coolest core of the PNe discussed in this paper. Nevertheless, BD+30 is a luminous object; at 10$^{37}$ ergs s$^{-1}$, its CSPN UV luminosity is rivaled only by NGC 7027. It is also among the brightest diffuse PN X-ray sources \citep{Kastner12}, and has been studied extensively across the electromagnetic spectrum \citep{Freeman16}. Still, it remains poorly characterized in molecular emission.

The structure of BD+30 is that of an elliptical nebula, with asymmetric CO bullets along possibly precessing jets \citep[and references therein]{Bachiller91, Akras11, Gussie95}. These misaligned jets may also be responsible for a hot bubble of X-ray-emitting material bounded by the ionized inner shell and outer dusty region \citep{Akras11, Freeman16}. An extensive multi-wavelength 3D structure modeling study was performed by \cite{Freeman16}.

In our survey data for BD+30, only HCO$^+$ was detected (Figure \ref{stackedBD+}). BD+30 subtends a diameter of 4" \citep{Frew08} and its CO emission is compact \citep{Bachiller91}; hence, beam dilution may have a deleterious effect on CO and other molecules surveyed. The HCO$^+$ line profile displays a fairly broad structure, with two distinct peaks. % whose velocities roughly correspond to those of the CO bullets \citep{Bachiller91}. 
We find the expansion velocity of these regions within BD+30 to be -46.0 km s$^{-1}$ and 38.4 km s$^{-1}$ respectively. Compared with the observations by \cite{Bachiller91}, who determined CO bullet velocities of -63 km s$^{-1}$ and 41 km s$^{-1}$, a puzzling discrepancy emerges. The redshifted HCO$^+$ component matches the CO velocity, but the approaching velocity component differs by $\sim$17 km s$^{-1}$, a significant offset. The velocity extension of the HCO$^+$ emission suggests that the molecular ion is not contained solely within the CO bullets but spread across the PN. Future followup with interferometer observations will be necessary to confirm this interference.

As in the case of NGC 7027, examination of the spectra revealed the presence of strong H41$\alpha$ and H38$\beta$ emission lines (Figure \ref{Hydrogen}), with peak intensities 0.076 K, 0.033 K and integrated flux values of 1.02 K km s$^{-1}$, 1.02 K km s$^{-1}$, respectively. These H recombination lines will not be further discussed in this paper, but they are consistent with the presence of a large mass of ionized hydrogen in BD+30.

\subsubsection{NGC 6445} %%%%% NGC 6445 %%%%%
\label{6445}

	%Figure of 6445 stacked molecular lines
	\begin{figure}
		\centering
		\resizebox{\hsize}{!}{\includegraphics[trim={7cm 2.5cm 8cm 2cm}, clip, width=0.7\textwidth]{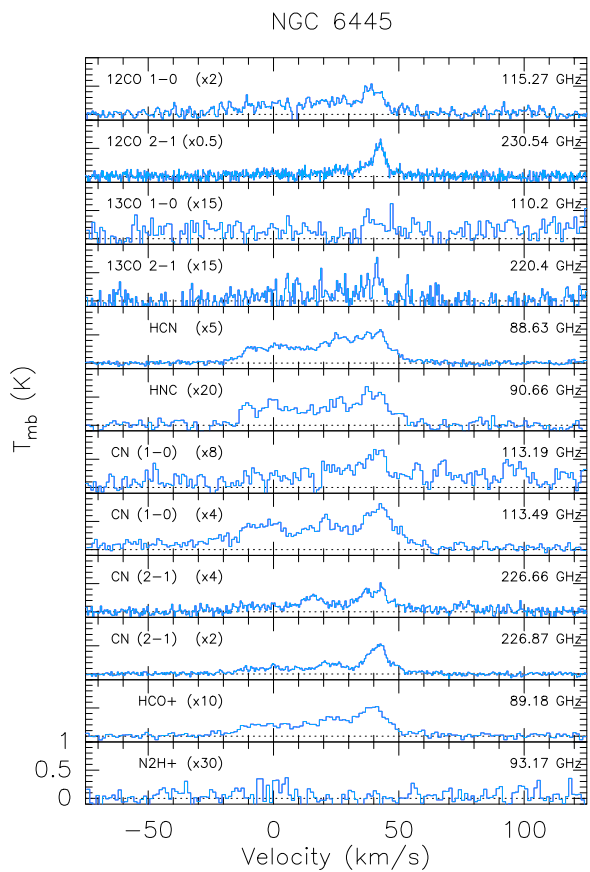}}
		\caption{As in Figure \ref{stacked7027}, molecular transitions detected and tentatively detected in the survey data for NGC 6445.}
		\label{stacked6445}
	\end{figure}
	
\indent NGC 6445, or the Little Gem, is the most distant PN in this survey, at 1.38 kpc \citep{Frew16}. It appears as a bipolar nebula of dynamical age roughly 5,000 years \citep{Phillips84} and its CSPN has a temperature of $T_\star$=170,000 K. NGC 6445 is at a comparable effective temperature and CSPN mass to NGC 7027, though it has far lower CSPN luminosity and larger nebula as a consequence of its more evolved state. The ChanPlaNs survey established that NGC 6445 harbors a compact X-ray source of uncertain physical origin, with coronal emission from a binary companion being a likely source \citep{Montez15}. The CO isotopologues along with OH and OH$^+$ were the only molecules previously detected in NGC 6445 \citep{Huggins89, Sun00, Aleman14}.

Here, we present the detection of five new molecular species: $^{13}$CO, HCO$^+$, CN, HCN, and HNC.
The shapes of these emission lines vary, with the nitrogen-bearing molecules displaying the broadest profiles (Figure \ref{stacked6445}). 
For those lines that are well-detected, double peaked structure is apparent and provides an estimated systemic velocity for the PN of 18.6 km s$^{-1}$ and outflow velocity of 22.0 km s$^{-1}$. The lack of a blueshifted edge in the CO lines suggests weaker CO emission from the approaching side of NGC 6445. Nevertheless, these velocity measurements are comparable with those of \cite{Sun00} who measured a CO (1-0) systemic velocity of 20 km s$^{-1}$ and an expansion velocity of 33 km s$^{-1}$. A feature near the systemic velocity at 20.4 km s$^{-1}$ can also be seen in the HCO$^+$ and CN lines. The $^{12}$CO (1-0) transition also shows a dip in its emission line spectrum at 115.276 GHz that is likely the result of over-subtracted ISM molecular emission along the line of sight. There is some hint of N$_2$H$^+$ emission at $\sim$0 km s$^{-1}$, within the velocity range of the other well-detected emission lines, but given the absence of emission at $\sim$40 km s$^{-1}$ -- where all the other lines peak -- we do not regard this as a detection of N$_2$H$^+$.

\subsubsection{NGC 7008} %%%%% NGC 7008 %%%%%
\label{7008}

        	\begin{figure}
        		\centering
        		\resizebox{\hsize}{!}{\includegraphics[trim={2cm 2.5cm 3cm 2cm}, clip, width=0.7\textwidth]{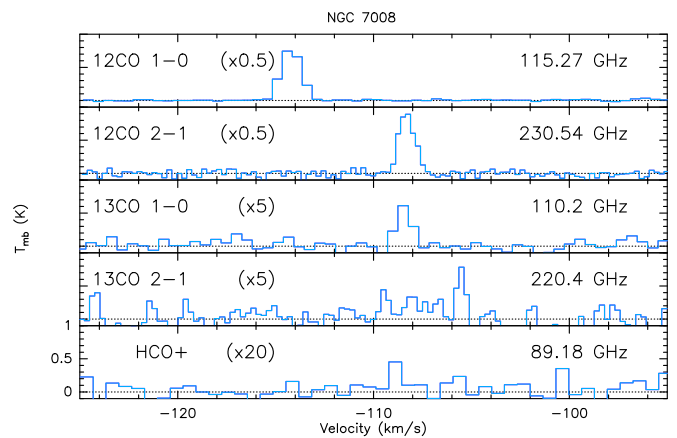}}
        		\caption{As in Figure \ref{stacked7027}, molecular transitions tentatively detected in the survey data for NGC 7008.}
        		\label{zoom7008}
        	\end{figure}
	
\indent NGC 7008 is a relatively young nebula, with a dynamical age of 5,700 years \citep{Schmidt16}. Despite not being previously detected in molecular line emission, it was included in this survey for its strong X-ray emission; as in the case of NGC 6445, the CSPN is detected as a point-like X-ray source with an unconstrained emission origin \citep{Montez15}. Previous molecular line measurements have been restricted to upper limits placed on HCO$^+$ and HCN \citep{Schmidt16}.

Observations were made along the northern edge of the molecular shell of NGC 7008 (Figure \ref{PNePositions}), yielding detections of narrow CO lines. 
Given the narrow widths and varying systemic velocities of the detected lines (Figure \ref{zoom7008}), we believe these emission lines of CO isotopologues originate from the ISM rather than from the nebula. Followup observations of NGC 7008 are necessary to establish the origin of CO emission towards the PN.

\subsubsection{NGC 6853} %%%%% NGC 6853 %%%%%
\label{6853}

        %Figure of 6853 stacked molecular lines
        	\begin{figure}
        		\centering
        		\resizebox{\hsize}{!}{\includegraphics[trim={6cm 2.5cm 6cm 1cm}, clip, width=0.7\textwidth]{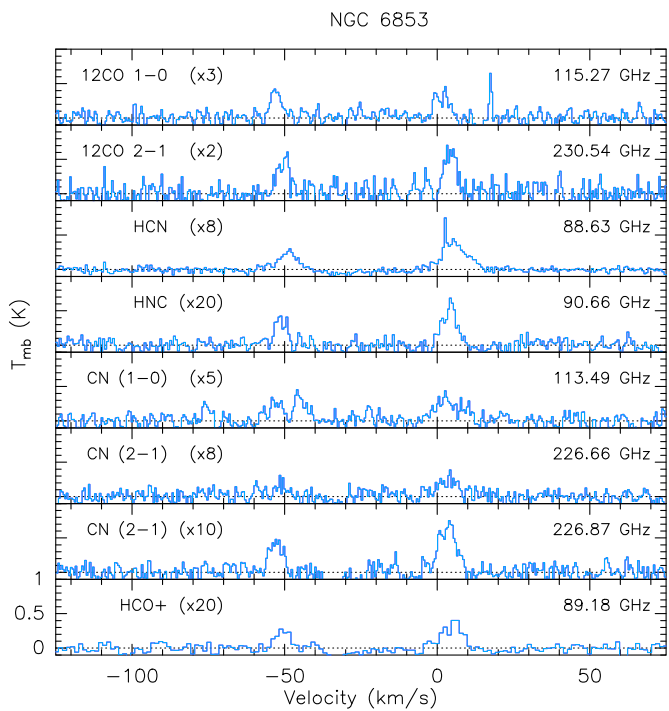}}
        		\caption{As in Figure \ref{stacked7027}, molecular transitions detected in the survey data for NGC 6853.}
        		\label{stacked6853}
        	\end{figure}

\indent NGC 6853, also known as M27 or the Dumbbell Nebula, is a large PN at 0.74 pc in diameter that subtends 7,500 arcsec$^2$ on the sky. In near-IR H$_2$ emission, the Dumbbell Nebula consists of dense knots and `streamers' that appear to radiate out from the CSPN \citep{Kastner96, Bachiller2000}. Numerous atomic lines in the optical such as [O I] and [N II] further trace out the eponymous `dumbbell' shape that reflect the PN's bipolar structure \citep{Edwards14}. In the ChanPlaNs survey, NGC 6853 was detected as a point-like source of X-ray emission, with intensity and low median energy as expected from photospheric emission due to the hot ($T_\star$=135,000 K) CSPN, given its small distance of 0.43 kpc \citep{Frew16}.

CO was first detected in NGC 6853 by \cite{Huggins96}, who identified a complex filamentary structure much like that seen in H$_2$. Further observations were carried out by \cite{Bachiller2000} and more recently by \cite{Edwards14}. In the latter, the study focused on a position offset from the CSPN ($\Delta\alpha$=-68", $\Delta\gamma$=-63"), where transitions of CO, HCO$^+$, and CS were identified. The region of the nebula targeted here is centered on the CSPN (Figure \ref{PNePositions}). H$_2$ imaging revealed this line-of-sight intersects a large column of clumpy, molecular gas \citep{Kastner96}.

In our 30 m spectra, a well defined double-peaked structure is observed, with blueshifted and redshifted components of -50.3 km s$^{-1}$ and 4.9 km s$^{-1}$, respectively (Figure \ref{stacked6853}). When compared with the velocities of emission line components detected by \cite{Edwards14} towards the edge of the PN (-35 km s$^{-1}$ and -5 km s$^{-1}$, respectively) the velocity components we detect hint at the 3D structure of the PN. The measured velocity components indicate a systemic velocity of -22.7 km s$^{-1}$ and expansion velocity of 27.6 km s$^{-1}$ for the expanding molecular shell of NGC 6853.

\subsubsection{NGC 6772} %%%%% NGC 6772 %%%%%
\label{endPN}

	%Figure of 6772 stacked molecular lines
		\begin{figure}
			\centering
			\resizebox{\hsize}{!}{\includegraphics[trim={6.5cm 2.5cm 7cm 2cm}, clip, width=0.7\textwidth]{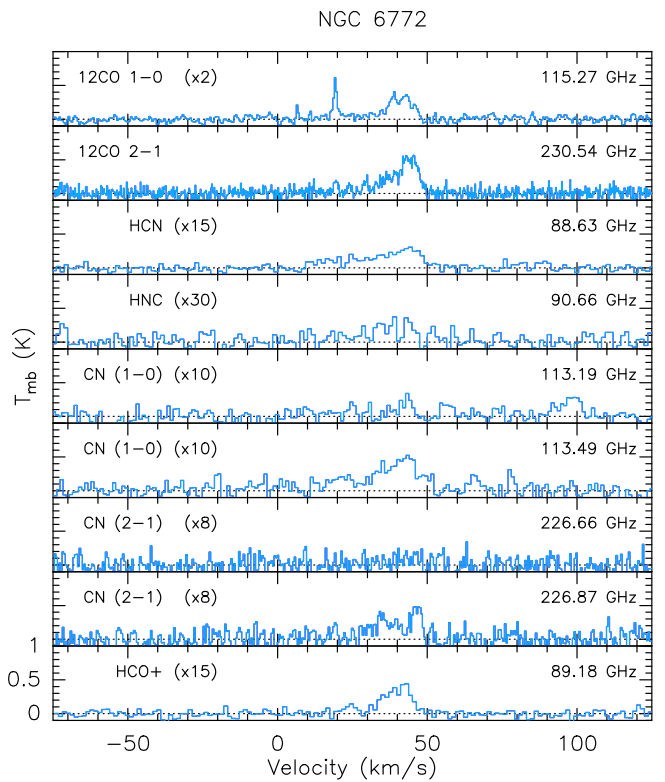}}
			\caption{As in Figure \ref{stacked7027}, molecular transitions detected in the survey data for NGC 6772.}
			\label{stacked6772}
		\end{figure}
		
\indent One of the older objects in this sample, with a dynamical age of $\sim$11,000 years \citep{Ali12}, the morphology of NGC 6772 is that of an elliptical shell with apparent distortion along the NE to SW axis. This structure may be a result of interactions between the expanding shell and the ISM \citep{Schmidt16}. NGC 6772 is one of three PNe in this paper not detected in X-rays in the ChanPlaNS survey \citep{Kastner12}. First detected in CO (1-0) by \cite{Zuckerman90}, \cite{Schmidt16} have more recently identified emission lines of CO (2-1) and HCN.

Our molecular line data were collected along the southern edge of the envelope (Figure \ref{PNePositions}). The 30 m spectra yield detections of both previously detected CO transitions, as well as $^{13}$CO (2-1), CN, and HCO$^+$ (Figure \ref{stacked6772}). The broad peaks of the spectral line profiles terminate at a redshifted velocity of approximately 44.2 km s$^{-1}$. This is consistent with the value determined by \cite{Schmidt16}. Only $^{12}$CO (1-0) displays a narrow blueshifted peak at roughly 19.3 km s$^{-1}$, as well as an apparent narrow feature near 0 km s$^{-1}$ that was also identified by \cite{Schmidt16}. A possible explanation for this narrow feature is interstellar gas along the line of sight to NGC 6772.

%%%%%%%%%%%%%%% DISCUSSION %%%%%%%%%%%%%%%
\section{Discussion}
%Discuss what the HNC/HCN trend means and how useful it can be, while HCO$^+$ indicates X-rays not as obvious as thought

%\input{/Users/jtb1435/Documents/School_Work/Research/Observations/XPNE_Survey/Tables/Line_Intensities.tex} %Ratios
\begin{table}
	\begin{center}
	\caption{HNC/HCN Integrated Line Intensity Ratios}
	\label{Ratios}
	\begin{tabular}{l c r}
	\hline
	\hline
	Object & Integrated Line Intensity & Uncertainty \\
			& (K km s$^{-1}$)	&	\\
	\hline
	NGC 7027	&	0.027	& 0.004 \\
	NGC 6445	&	0.247	& 0.035 \\
	NGC 6720	&	0.366	& 0.052 \\
	NGC 6720 Rim	&	0.290	& 0.041 \\
	NGC 6853	&	0.320	& 0.045 \\
	NGC 6772	&	0.365	& 0.052 \\
	NGC 7293	&	0.717	& 0.101 \\
	NGC 6781	&	0.452	& 0.064 \\
	\hline
	\end{tabular}
	\end{center}
	\tablefoot{\\	Integrated line intensity values of the HNC/HCN ratio in observed PNe. 	}
\end{table}

\subsection{HNC/HCN Ratio}
The HNC/HCN abundance ratio has been studied for the insight it provides into the physical conditions within the photodissociation regions of cold molecular clouds. Activation energy barriers and the existence of a critical temperature for the conversion between HNC and HCN, both formed predominantly by dissociative recombination of HCNH$^+$, have long been discussed \citep{Schilke92, Bachiller97, Schmidt17}. For semi-ionized gas near the PDR boundary, heightened temperatures should favor production and survival of HCN, decreasing the HNC/HCN ratio. The lack of convergence on the fundamental mechanisms and parameters that establish the ratio is apparent in the current literature, however \citep[and references therein]{Graninger14}.

\cite{Schilke92} studied HNC/HCN as a diagnostic of local gas heating in the Orion Molecular Cloud (OMC-1), and in particular, explored the temperature dependence of the HNC+H$\rightarrow$HCN+H reaction. They found the HNC/HCN ratio to be $\sim$1 at temperatures of $\sim$10~K and noted that the ratio decreases in OMC-1 as the gas temperature increases from $\sim$10 to $\sim$150~K. \cite{Schilke92} then assumed activation energies for both the HNC+H and HNC+O reactions of $\sim$200~K in order to reproduce the observed ratios. Using gas-grain and gas-phase chemical modeling, \cite{Graninger14} furthered the study of HNC/HCN in OMC-1, where they identified the critical temperature range of the gas cloud at which the HNC/HCN ratio would begin to decrease from unity, i.e. 20-40~K, and placed upper limits on the activation barrier energy of well below 1200~K.

The conclusions from these foregoing observational studies are broadly consistent with those of \cite{Jin15}, who analyzed the HNC/HCN ratio across dark clouds, protostellar objects, and H{\sc ii} regions. \cite{Jin15} found the mean ratio decreased gradually across these evolutionary stages. They concluded that the HNC/HCN ratio can trace the evolutionary stages of massive star formation from cold molecular clouds ($\sim$20~K) to the warmer molecular gas ($\sim$100~K) associated with H{\sc ii} regions generated by young, massive (OB) stars. Similarly, the ring-like vs. centrally peaked morphologies of HNC and HCN, respectively, within the disk orbiting TW Hya appear to place constraints on the characteristic temperature necessary to efficiently convert HNC into HCN at $\sim$25 K \citep{Graninger15}.

The abundances of HNC and HCN in PNe can be expected to change due to thermal and kinetic processes as the envelope of AGB ejecta evolves to post-AGB and, with the CSPN reaching ionization temperatures, into PN stages \citep{Schmidt17}. As it is more stable, HCN is favored in the AGB stage where LTE chemistry dominates the inner envelope of the star \citep{Schmidt17}. \cite{Herpin02} also observe lower abundances of  HNC in post-AGB stages and claim this is due to increased ion-molecule interactions. By the proto-PN phase, abundances of HNC and HCN are expected to restabilize to unity \citep{Schmidt17, Kimura12}. In more evolved PNe, models indicate the increased ionization fractions should result in enhanced abundances of HNC relative to HCN \citep{Bachiller97}.

\subsubsection{HNC/HCN Ratio with CSPN L$_{UV}$}
\label{HNC/HCN}
Eight PNe were observed in both the HNC and HCN during our molecular line survey. We find that the observed HNC/HCN line intensity ratio ranged from 0.03 to 0.72 within the objects sampled (Table \ref{Ratios}).  

\begin{figure}
		\centering
		\resizebox{\hsize}{!}{\includegraphics[trim={2cm 0cm 4cm 2cm}, clip, width=0.8\textwidth]{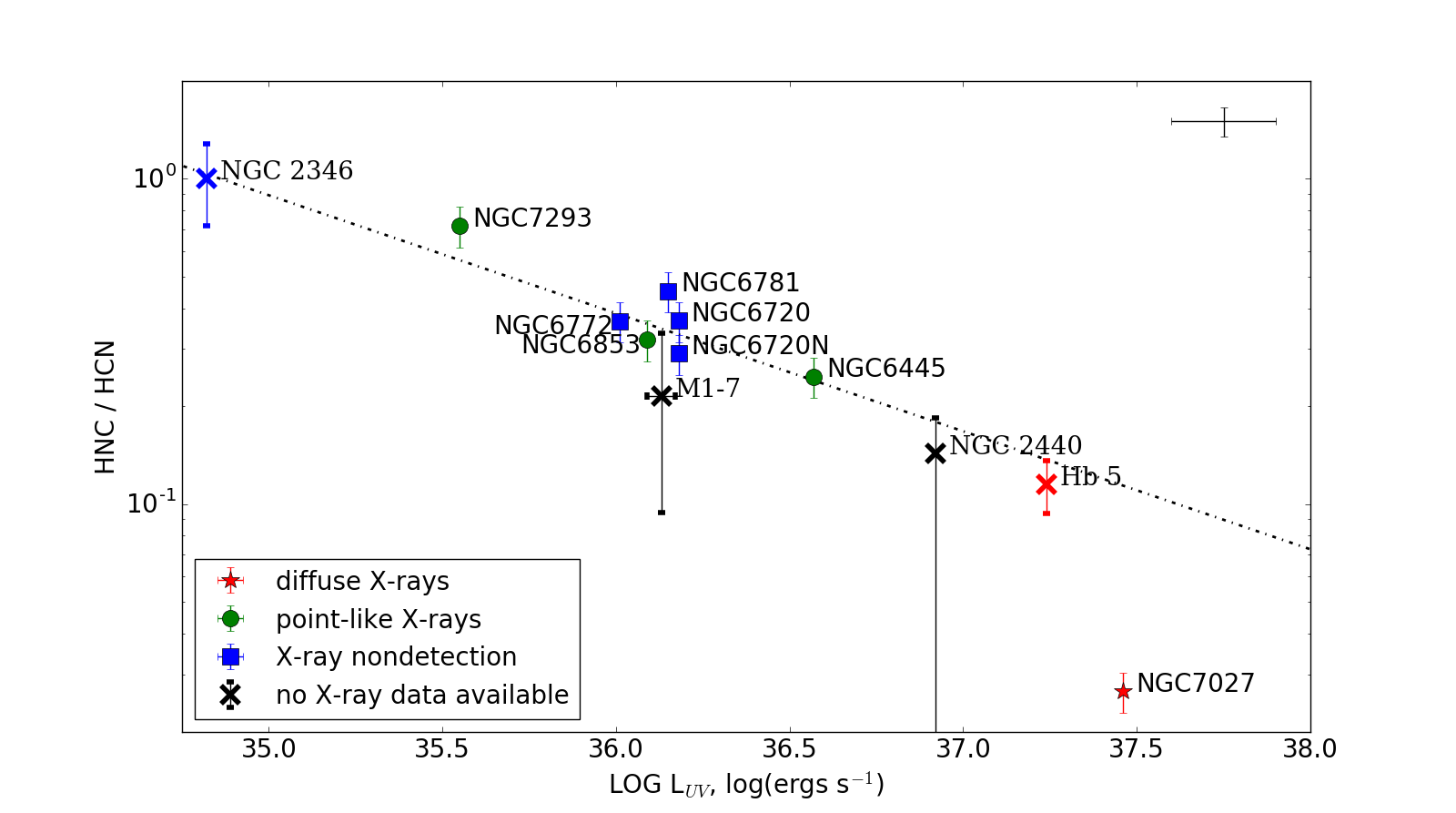}}
		\caption{Flux ratio of HNC to HCN plotted against PN central star UV luminosity. Line ratios obtained from observations by \cite{Schmidt17} are denoted by crosses. CSPN UV luminosities from Table \ref{T1A1} and \citet[NGC 2346]{Bachiller89+}, \citet[NGC 2440 and Hb 5]{Frew08}, and \citet[M1-7]{Sabbadin84}. The best-fit linear power-law slope (black dashed line, NGC 7027 and literature sources omitted) is $m=-0.363$ with a linear correlation coefficient of $r=-0.885$. Typical uncertainty estimate of 40\% for UV luminosity and 10\% for line ratios is plotted in the top right.}
		\label{HNC_HCN_UV}
	\end{figure}

Notably, this analysis has revealed a previously unknown correlation between the HNC/HCN line intensity ratio and the UV luminosity of the central star. In Figure \ref{HNC_HCN_UV}, we plot the HNC/HCN ratios obtained from our 30 m data, as well as data from other nearby PNe taken from the literature, versus CSPN UV luminosity. The HNC/HCN line ratios for the PNe sample decrease from roughly unity to 0.027, as the CSPN UV luminosity increases. Linear regression for the data for our survey objects (omitting NGC 7027) supports the presence of a power-law relationship between the HNC/HCN ratio and CSPN UV luminosity, with a best-fit slope of $m=-0.363$ and a correlation coefficient of $r=-0.885$. There exists no dependence of the HNC/HCN ratio on the type or presence of an X-ray source.

For the HNC/HCN line ratio to decrease steadily with UV luminosity, high energy photons must be either heating the gas within the PDR or selectively photodissociating HNC \citep{Aguado17}. Under the former scenario, as CSPN UV luminosity rises, the resulting CSPN-induced ionization evidently raises the local temperature to exceed the activation energy for conversion of HNC into HCN+H. 
\cite{Graninger15} suggest that when the local PDR is heated to the range of 25-40~K, the HNC+H reaction will lead to a decreased HNC/HCN ratio. Under this scenario, PDR heating would therefore be a direct result of photoelectrons generated by UV CSPN emission. 
Recent observations targeting proto-brown dwarfs observed a similar decrease in the HNC/HCN abundance ratio with bolometric luminosity for the case of protostars \citep{Riaz18}. \cite{Riaz18} suggest that this trend is due to the increased efficiency of heating of molecular gas, resulting in declining HNC abundance with increasing protostellar luminosity.

Another potential mechanism that might drive the HNC/HCN ratio is the difference in photodissociation rates of HNC and HCN. Calculations by \cite{Aguado17}, taking into account the photodissociation cross sections of each molecule as well as variation in the incident UV radiation field, find that selective photodissociation can increase the rate of destruction of HNC by a factor of $\sim$2-10 relative to that of HCN. While high effective temperature sources, such as CSPNe, have fewer low-energy photons and thus a lower effect on photodissociation rates than cool sources, the interaction between the strength of the evolving UV source and the photochemical reaction rates of HNC and HCN could be sufficient to drive the observed trend.

Modeling of HNC and HCN generation in cold, UV-irradiated molecular regions is necessary to identify the relevant processes that are most important in determining HNC/HCN variation in molecule-rich PNe. 
In addition, having established the correlation of HNC/HCN ratio with UV luminosity for an ensemble of PNe, it is necessary to investigate whether and how the HNC/HCN ratio depends on the local UV flux within individual objects. 

Note that, since excitation of HCN and HNC is similar under similar physical conditions, we expect the trend in their line intensity ratio apparent in Figure \ref{HNC_HCN_UV} (and the weaker trend in Figure \ref{Age}; see Section \ref{RatioAge}) to closely track a similar trend in their column density ratio.

%Figure of HNC to HCN fraction against Luv

\subsubsection{HNC/HCN Ratio with Nebular Age}
\label{RatioAge}

\begin{figure}
		\centering
		\resizebox{\hsize}{!}{\includegraphics[trim={2cm 0cm 3cm 1cm}, clip, width=0.8\textwidth]{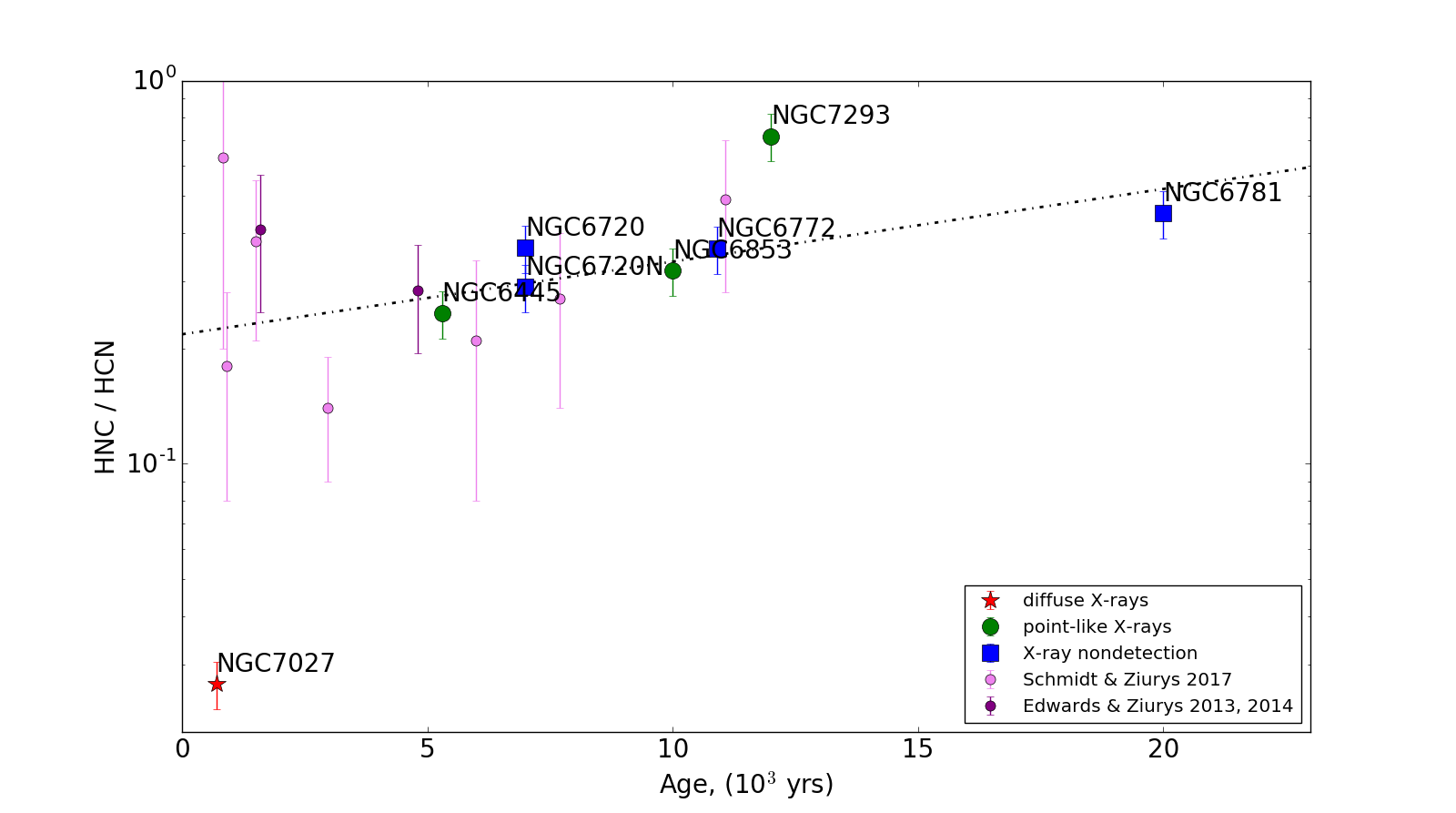}}
		\caption{Flux ratio of HNC/HCN plotted against nebular age. Magenta points are PNe observed by \cite{Schmidt17}. Their sample encompasses objects with an age range half that of the PN included in our study. Linear regression identified a correlation of $r=0.496$ with a linear slope of $m=0.019$, or an increase in the ratio of HNC/HCN of 0.019/kyr.}
		\label{Age}
	\end{figure}

In the past, \cite{Herpin02} suggested that the emergence of a strong UV source early in a PN's history generates a significant drop in HCN with respect to HNC through PDR processing. Recent work on the HNC/HCN ratio in young PNe found that there is insignificant change with dynamical age \citep{Schmidt17}. These results were taken to indicate that the ratio does not vary significantly with nebular evolutionary state, and after the proto-PN stage, remains fixed. Because our survey samples PNe spanning a somewhat larger range of dynamical age, we have revisited the question of whether HNC/HCN correlates with PN age (Figure \ref{Age}). Linear regression to our 30 m survey data (omitting NGC 7027) yields a correlation coefficient $r=0.496$, suggesting a marginal correlation. When including data from the literature (pink and purple dots in Figure \ref{Age}) \citep{Schmidt17, EdwardsZ14, Edwards13}, however, we find no correlation between HNC/HCN ratio and PN age. Nevertheless, the trend obtained from our data appears to match reasonably well with these previous PN survey data. These results support the notion that a continued rise in HNC with respect to HCN occurs throughout PN evolution. Due to the nature of PN age estimates, however, uncertainties in the ages are difficult to define and thus have not been included in the plot. Additional HNC and HCN observations of evolved PNe are necessary to confirm the tentative trend indicated in Figure \ref{Age}.

%Figure of HCN/HNC fraction against PN Age

\subsection{Diagnostic of X-irradiated Gas}
%Lack of correlation with X-rays
\label{HCO+}

We investigated several potential tracers of X-ray-induced chemistry among the molecular emission lines. Line ratios were compared against X-ray and UV luminosity of each central star, as well as the ratio of the X-ray and UV luminosities. No significant trends emerged from this analysis. We focus here on HCO$^+$, which has long been proposed as a diagnostic of X-irradiation of molecular gas \citep{Deguchi90}. 
Specifically, HCO$^+$ is generated via X-ray ionization of molecular gas through the reaction CO$^+$+H$_2$$\rightarrow$HCO$^+$+H or by the molecular ion H$_3^+$ through the reaction H$_3^+$+CO$\rightarrow$HCO$^+$+H$_2$ \citep{Zhang08}. An enhanced abundance of HCO$^+$ may then act as a tracer for X-ray emission in the PNe, and we would expect line ratios correlated with CSPN or nebular X-ray luminosity.

\begin{figure}[h]
		\centering
		\resizebox{\hsize}{!}{\includegraphics[trim={2cm 0cm 2cm 1cm}, clip, width=\textwidth]{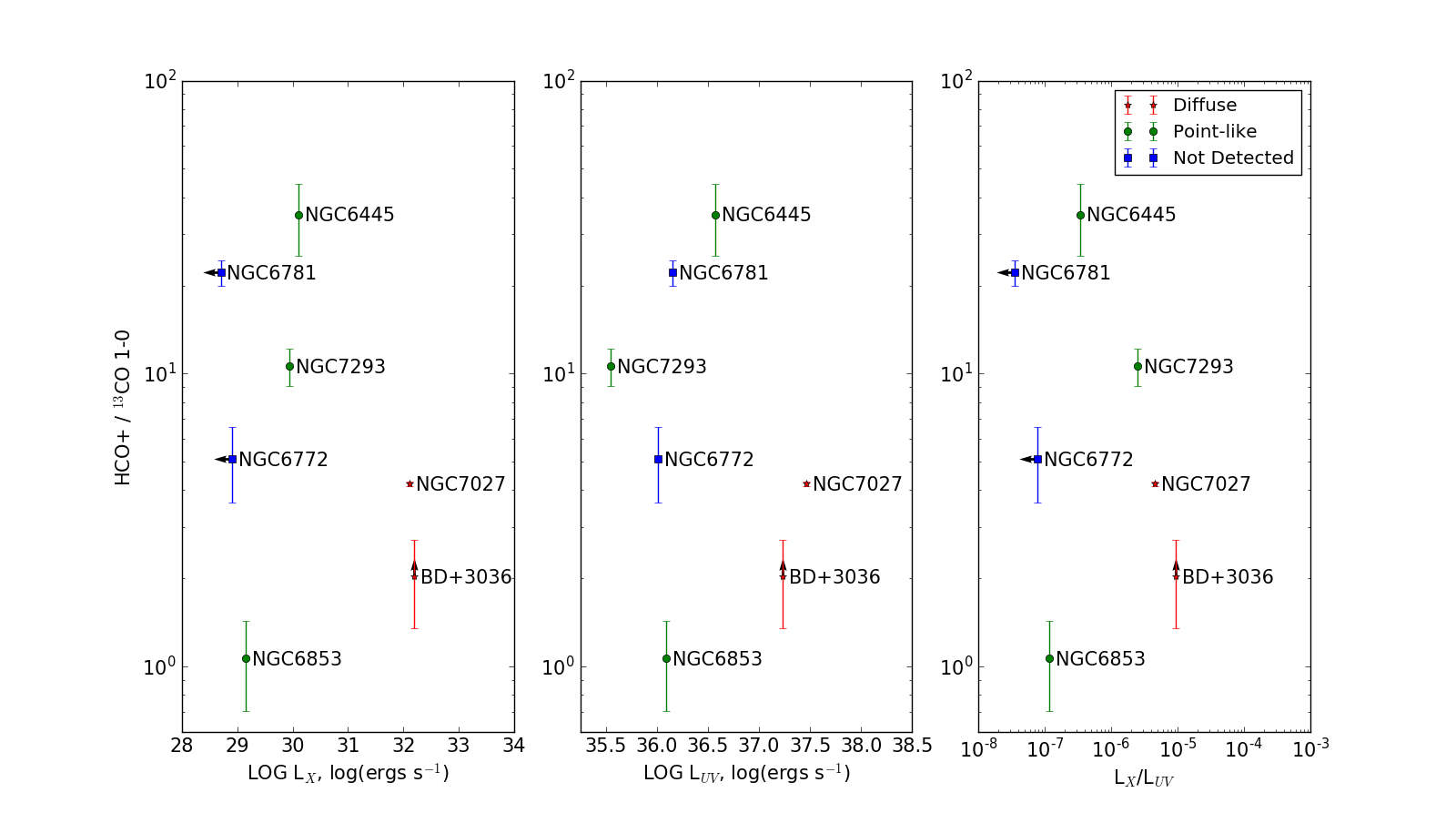}}
		\caption{HCO$^+$ to $^{13}$CO 1-0 line ratio plotted against the central star's luminosities L$_X$, L$_{UV}$, and L$_X$/L$_{UV}$. Arrows indicate upper limits.}
		\label{HCO_L}
	\end{figure}

Figure \ref{HCO_L} plots the ratio of HCO$^+$ to $^{13}$CO against CSPN UV and X-ray emission. $^{13}$CO was chosen as it acts as a proxy of the molecular mass and is not coupled to X-irradiation. The absence of a clear trend in the left-hand panel of Figure \ref{HCO_L} may suggest that X-rays are not directly driving HCO$^+$ production in PNe. On the other hand, Figure \ref{HCO_L} (center panel) also does not reveal a clear trend of enhanced HCO$^+$ in the presence of strong UV from CSPNe. Hence, it remains to establish the irradiation source that is primarily responsible for the production of this molecular ion in PNe.

%Figure of HCO$^+$ to 13 CO 1-0 fraction against Lx, Luv, and Lx/Luv

\subsection{CO$^+$ and N$_2$H$^+$}

In PNe with hot central stars or high energy emission, X-rays have the potential to drive the chemical pathways available to the molecular gas. CO$^+$ is produced in higher abundances in the presence of X-ray dominated regions through the reaction C$^+$+OH$\rightarrow$CO$^+$+H \citep{Spaans07}. An enhanced abundance of CO$^+$ then might suggest formation through this X-ray driven reaction. The main chemical pathway for the production of N$_2$H$^+$, i.e., H$_3^+$+N$_2$$\rightarrow$N$_2$H$^+$+H$_2$, also requires X-ray or cosmic ray irradiation as a means to generate the requisite H$_3^+$ ion. \citep{Zhang08, Bell07}.

Of the nine observed PNe, transitions of CO$^+$ and N$_2$H$^+$ were detected only in the spectra of NGC 7027. None of the other objects surveyed here have had confirmed detections of CO$^+$, except NGC 6781, despite its potential importance in the generation of HCO$^+$ \citep{Bell07}. Considering the CO$^+$ line was quite weak in NGC 7027 compared to $^{12}$CO, the molecule may be present in other X-ray emitting PNe, but at a level below our survey sensitivity. Given that NGC 7027 harbors an unusually luminous diffuse X-ray source and an unusually large mass of molecular gas, these detections hint at the viability of CO$^+$ and N$_2$H$^+$ diagnostics of X-irradiation of molecular gas. 
Shocks may also play a role in N$_2$H$^+$ formation, however, and future observations are required to distinguish these two mechanisms. %Previous works have also shown the utility of ionized molecules for diagnosing X-irradiated gas \citep{Spaans07, Zhang08}.

%%%%%%%%%%%%%%% CONCLUSIONS %%%%%%%%%%%%%%%
\section{Summary}

This study presents observations of key molecular transitions across the 88-236 GHz range toward nine planetary nebulae with the IRAM 30 m radio telescope. We report new detections for five of the PNe in the survey across thirteen molecular lines included in our spectral coverage. 
Our survey yielded new detections of molecular transitions of CO, CN, HCN, HNC, and/or HCO$^+$ in NGC 6445, NGC 6853, and NGC 6772. The molecular ion HCO$^+$ was also detected in BD+30$\degree$3639 for the first time.  
In addition, confirmation of previously detected molecules were made in NGC 7027, NGC 6720, NGC 7293, and NGC 6781. 

Our analysis of line ratios that are potential diagnostics of high-energy irradiation has revealed a strong but previously unrecognized anticorrelation between the HNC/HCN line intensity ratio and CSPN UV luminosity. This anticorrelation, which persists over nearly 3 orders of magnitude in central star UV luminosity, provides strong evidence that HNC/HCN acts as a tracer for heating of the nebular molecular gas by UV photons.
Our survey additionally found a marginal correlation between the HNC/HCN line ratio and PN age.

Other irradiation tracers explored in our PN molecular line survey yield more ambiguous results. Though predicted to increase in abundance with X-irradiation, we find the intensity of HCO$^+$ line emission in PNe shows no clear correlation with PN X-ray luminosity. Detections of CO$^+$ and N$_2$H$^+$ in our survey were limited to the particularly molecule-rich and luminous NGC 7027, which had been previously detected in lines of both species. This suggests that CO$^+$ and N$_2$H$^+$ trace especially intense high-energy irradiation of molecular gas.

%\vspace{1em}

\begin{acknowledgements}
This work was supported in part by a subcontract issued to RIT under NASA ADAP grant \#80NSSC17K0057 to STScI (PI: B. Sargent), and by the Spanish MINECO within the program AYA2016-78994-P. It is also based on observations carried out with the IRAM telescopes. IRAM is supported by INSU/CNRS (France), MPG (Germany) and IGN (Spain). JB wishes to acknowledge useful discussions with Pierre Hily-Blant and Thierry Forveille during his Fall 2018 residency at IPAG under the support of the Chateaubriand Fellowship of the Office for Science \& Technology of the Embassy of France in the United States.
\end{acknowledgements}

%%%%%%%%%%%%%%% REFERENCES %%%%%%%%%%%%%%%
\bibliographystyle{apj}
\bibliography{High_Energy_Ref.bib}

\end{document}